\documentclass[fleqn,usenatbib]{mnras}

\newcommand{\et}{{et al.\ }}

\newcommand\cdof{{\rm C/{\rm dof}}}

\newcommand\dec{{\Delta {\rm C}}}

\newcommand\oiii{{[O\,\textsc{iii}]}}
\newcommand\ovii{{O\,\textsc{vii}}}
\newcommand\oviii{{O\,\textsc{viii}}}
\newcommand\neix{{Ne\,\textsc{ix}}}

\newcommand\fexvii{{Fe\,\textsc{xvii}}}

\newcommand\halpha{{H$\alpha$}}
\newcommand\hbeta{{H$\beta$}}
\newcommand\NeOx{{Ne/O}}
\newcommand\NOx{{N/O}}
\newcommand{\nh}{{$N_{\rm H}$}}
\newcommand\cts{{\rm\thinspace count}}
\newcommand\cps{\hbox{$\cts\s^{-1}\,$}}

\newcommand\cm{{\rm\thinspace cm}}

\newcommand\erg{{\rm\thinspace erg}}

\newcommand\kev{{\rm\thinspace keV}}

\newcommand\anorm{photons/cm$^2$/s}

\newcommand\pc{{\rm\thinspace pc}}
\newcommand\Msun{\hbox{$\rm\thinspace M_{\odot}$}}

\newcommand\ks{{\rm\thinspace ks}}
\newcommand\s{{\rm\thinspace s}}

\newcommand\ergps{\hbox{$\erg\s^{-1}\,$}}

\newcommand{\angstrom}{\mbox{\normalfont\AA}}
\newcommand{\mekal}{{\sc mekal}}
\newcommand{\apec}{{\sc apec}}
\newcommand{\pion}{{\sc pion}}
\newcommand{\xstar}{{\sc xstar}}
\newcommand{\chandra}{{\it Chandra}}

\newcommand{\rosat}{{\it ROSAT}}

\newcommand{\suzaku}{{\it Suzaku}}

\newcommand{\xmm}{{\it XMM-Newton}}

\newcommand{\nustar}{\textit{NuSTAR}}

\newcommand{\xspec}{XSPEC}

\newcommand{\kms}{km~s$^{-1}$}

\newcommand{\mrk}{Mrk\thinspace1239}

\usepackage{newtxtext,newtxmath}
\usepackage[T1]{fontenc}
\usepackage{float}

\usepackage{graphicx}	% Including figure files
\usepackage{amsmath}	% Advanced maths commands
\usepackage{siunitx}

\title[Mrk 1239 - RGS]{The Collisional and Photoionized Plasma in the Polarized NLS1 galaxy Mrk\,1239}

\author[M. Z. Buhariwalla et al.]{
Margaret Z. Buhariwalla,$^{1}$\thanks{E-mail: margaret.buhariwalla@smu.ca}
Luigi C. Gallo,$^{1}$
J. Mao,$^{2,3}$
S. Komossa,$^{4}$
J. Jiang,$^{5}$
A Gonzalez,$^{1}$
and D. Grupe$^{6}$
\\
$^{1}$Department of Astronomy and Physics, Saint Mary's University, 923 Robie Street, Halifax, NS B3H 3C3, Canada\\
$^{2}$Department of Astronomy, Tsinghua Univerisity, 30 Shuangqing Road, Beĳing 100084, China\\
$^{3}$Department of Physics, Hiroshima University, 1-3-1 Kagamiyama, Higashi-Hiroshima City, Hiroshima 739-8526, Japan \\
$^{4}$Max-Planck-Institut f{\"u}r Radioastronomie, Auf dem H{\"u}gel 69, 53121, Bonn Germany \\
$^{5}$Institute of Astronomy, University of Cambridge, Madingley Road, Cambridge CB3 0HA UK \\
$^{6}$Department of Physics, Geology, and Engineering Technology Science Center, Northern Kentucky University, 1 Nunn Drive, Highland Heights, KY 41099
}
\date{Accepted XXX. Received YYY; in original form ZZZ}

\pubyear{2023}

\begin{document}
\label{firstpage}
\pagerange{\pageref{firstpage}--\pageref{lastpage}}
\maketitle
\begin{abstract}
\mrk\ is a highly  polarized NLS1 in the optical band, whose $0.3-3$\kev\ spectrum has remained remarkably consistent  over more than two decades of observation. Previous analysis of this object suggested that the soft X-ray band was dominated by emission lines (collisionally and/or  photoionized) from the distant host galaxy as the X-ray emission from the central engine was highly obscured.  New \xmm\ data of \mrk\ are presented here to investigate the soft X-ray band of this galaxy with high resolution. The first RGS spectra of this source reveal a plethora of ionized emission lines  originating from two distinct plasmas, one collisionally ionized and the other photoionized at approximately equal brightness. The best fit model uses \apec\ and \xstar\ grids to account for the collisionally ionized and photoionized components, respectively. The fit improves significantly if the photoionized material is allowed to outflow at $\approx 500$ \kms, matching the outflow velocity of the  forbidden \ovii\ emission line.  From constraints on the ionization and density of the photoionized material we can estimate the location of it to be  no further than a few pc from the central source, around the outer radius of the torus, which is consistent with  the \ovii$(f)$ emission line.   Properties of the collisionally ionized  plasma are consistent with star formation  rate (SFR) of  $\approx 3 \Msun  \textrm{yr}^{-1}$, which is comparable with several previous measurements of the SFR in  this galaxy.
\end{abstract}


%%%%%%%%%%%%%%%%%%%%%%%%%%%%%%%%%%%%%%%%%%%%%%%%%%

%%%%%%%%%%%%%%%%% BODY OF PAPER %%%%%%%%%%%%%%%%%%

\section{Introduction}
\mrk\ is a Narrow Line Seyfert 1  galaxy (NLS1) located at a redshift of $z=0.01993$ \citep{Beers+1995}. The optical spectra of \mrk\ was first reported by \cite{Rafanelli+1984} and later used by \cite{Osterbrock+1985} to define the subclass of galaxies known as NLS1s. They suggest a classification criterion for NLS1 to have  \hbeta\ line widths less than 2000\,\kms. This is now commonly adopted by the community \cite[see][for a complete review]{Komossa+2008}. These active galaxies are thought to be low mass counterparts of broad line Seyfert 1s  (BLS1). They show intense X-ray variability on both long-term and rapid time scales (see \citealt{Gallo+2018,Wilkins+2017, Wilkins+2015B}). These objects are seen through low column densities and often exhibit steep powerlaw continuum \citep{Waddell+2020} and strong soft excesses \citep{Waddell+2019}.

\mrk\ is a radio quite galaxy that shows evidence of non-stellar radio emission \citep{Doi+2015}. This indicates the possible presence of a radio jet originating from the AGN in this galaxy. \mrk\ is also a highly polarized in the broad \halpha\ and \hbeta\ emission lines, along with the continuum \citep{Pan+2019, Pan+2021}. The high levels of polarization in \mrk\ have previously been observed by \cite{Goodrich+1989} where \mrk\ was the highest polarized source in the sample. This suggests some amount of scattered emission in this source. \mrk\ is classified as a polar-scattered Seyfert 1 galaxy, meaning the line of sight is through the upper layers of the tours \citep{Smith+2004,Jiang+2019}.
As well \cite{Goodrich+1989} found that the Balmer lines had a higher polarization than the forbidden \oiii\ lines. This is strong evidence that the sources of these emissions lines is physically distinct. 

\mrk\ was first observed in  X-ray using \rosat\ where a steep photon index was found  \citep[$\Gamma \simeq 3$,][]{Rush+1996,Jiang+2021}.  \mrk\ was a serendipitous source in a 2001 observation by \xmm\ and analysed by \cite{Grupe+2004}. They found the continuum was best fit using a powerlaw with two intrinsic absorbers: the first a direct, highly absorbed path and the second a less absorbed scattered path. They concluded that this was consistent with the distinct emitting/scattering regions found in \cite{Goodrich+1989}.  Finally, \cite{Grupe+2004} found an excess of emission at 0.91 \kev, which was attributed to a blend of the \neix\ triplet.  However, the \ovii\ triplet was not observed in the spectrum suggesting a super-solar ratio of \NeOx\ in \mrk. 

\mrk\ was revisited by \cite{Buhariwalla+2020} with archival X-ray data from \suzaku\ and \nustar, obtained in the subsequent years since \cite{Grupe+2004}. With this broadband multi-epoch analysis  some truly unique behaviour  was revealed. First and foremost, across the 18 years of data analysed the shape and flux of the spectra below 3\,\kev\ remain remarkably similar. Meanwhile, the spectrum above 3\kev\ retained the characteristic NLS1 variability. This immediately indicated the presence of a strong absorbing feature that blotted out most of the emission from the central engine below 3\kev. The 0.91 \kev\ feature was present in the 2007 \suzaku\ observation of \mrk. 

Continuum modelling of this source between 0.3 and 30 \kev\ saw a ionized partial covering model compared to a blurred reflection model. The data were insufficient to significantly distinguish between the two models but their similarities proved to be more interesting. Both models required high levels of intrinsic absorption and both models required the presence of a collisionally ionized component to fit the spectrum below 3\kev. \cite{Buhariwalla+2020} attributed this collisionally ionized emission (CIE) to the presence of star forming activity in this object. They proposed that the 0.91 \kev\ feature found originally by \cite{Grupe+2004} was not an overabundance of Ne, but a blend of Fe-L lines from the regions of star formation. Using a relationship between the $L_{2-10\kev}$ of the CIE and star formation rate (SFR) \citep{Franceschini+2003}, the SFR of \mrk\ was found to be $\simeq4-6\Msun \, {\rm yr^{-1}}$. This is comparable with the SFR of $<7.5\Msun \, {\rm yr^{-1}}$measured using PAH signatures \citep{RuschelDutra+2017},  $\sim3.5\Msun \, {\rm yr^{-1}}$ found using SED fitting \citep{Gruppioni+2016}, and $\sim2.1\Msun \, {\rm yr^{-1}}$ found using IR measurements \citep{Smirnova+2022}.  To fully probe the soft band of this galaxy and determine the origin of the 0.91 \kev\ feature conclusively deeper observations of \mrk\ were needed. 

This paper presents the first high-resolution X-ray spectrum of \mrk\ in the $7-31$\,\AA\ band with the \xmm\  RGS spectra. Limited MOS/PN spectra of a deep observation of \mrk\ and a \chandra\ image is also presented. In Section \ref{sec:data}, the observations and data reduction techniques are summarized. Section \ref{sec:results} examines the \chandra\ image,  emission line identification,  possible outflow velocities, plasma diagnostics and the possibility of a \NeOx\  and in Section \ref{sec:spectral} the spectra are analyzed. A discussion of the results is given in Section \ref{sec:disc}, and conclusions are drawn in Section \ref{sec:conclusion}. 
\section{Observations and Data Reduction}
\label{sec:data}
A snapshot observation of \mrk\ was taken on 2021-04-11 with \chandra\ as part of a sample study exploring duel AGN \citep{Foord+2020}. Several months later a deep observation of \mrk\ was taken on 2021-11-04 with \xmm\ and \nustar, with each obtaining $\approx100$ \ks\ of on source time. During the second half of the \nustar\ observation \mrk\ entered a flaring state. To fully explore all facets of this object analysis of the \nustar\ and broad band \xmm\ data is relegated to a second paper.  The data for analysis are listed in  Table \ref{tab:obs}. This section describes the observations and data reduction.
\begin{table*}
	\begin{tabular}{c c c c c c c c}
		\hline
		(1) 				  & (2) 			 			& (3) 		 						  & (4) 		  				& (5) 	 			& (6) 				& (7) 	 		& (8)				\\
		Observatory & Observation ID   &  Instrument Name 	 & Start date  			& Duration 	& Exposure 		& Counts & Energy range		\\
								 & 				   				  & 			 							& (yyyy-mm-dd)	& [s]      			& [s]      				&    	 		& 			 		\\
		\hline
		\xmm			& 0891070101	  & PN 									 &2021-11-04 		 & 105000 	& 79175				& 1512	 &	$0.2-2.0$\,\kev	\\
		\xmm 	 		& 				    			  & RGS 		 					   &2021-11-04  		& 105000 & $193963^{\star}$	&$5660$	 &  $7-31$\,\AA 	\\
		\hline
		\chandra     & 23693		            & ACIS-S 							& 2021-04-11 		& 2100 		& 2021					& 50 	 & $0.2-2$ \kev  	\\ 
		\hline	
	\end{tabular}
	\caption{Observations log for Mrk 1239. The observations and  instruments used for analysis are listed in column (1). The observation ID and labels used in this work are given in columns (2) and (3), respectively. The start date of each observation is given in column (4). The duration of each observation, total exposure time and total counts for each observation are given in columns (5), (6), and  (7), respectively. The energy/wavelength each observation were fit over is given in column (8). $^{\star}$For RGS the combined exposure and counts for RGS1 and RGS2 are reported.}
	\label{tab:obs}
\end{table*}
\subsection{\xmm}
\mrk\ was observed with  \xmm\ \citep{Jansen+2001} for 105 \ks\ starting on November $4^{th}$ 2021. The \xmm\ Observation Data Files (ODF) were processed to produce calibrated event lists using the \xmm\ Science Analysis System, {\sc sas}   V17.0.0.
The first order RGS data was extracted from the ODF using {\sc rgsproc}. Upon analysis of the light curve a background flaring event was seen within the first 15 \ks. {\sc tabgtigen} was used to filter all instances where the \cps\ was greater than or equal to 0.1. The spectra was then re-extracted using {\sc rgsproc}. RGS1 and RGS2 were then combined using  {\sc rgscombine}, which produced a combined source spectra, background spectra, and response matrix. The source and background spectra were then optimally binned using {\sc ftgrouppha} \citep{optbin}, the spectra were loaded into {\sc xspec} where the background was modelled. Fits were evaluated  using Cash-statistics \citep[C-stat,][]{Cash+1979}.

For the PN spectra an event list was created from the ODF using {\sc epproc}. The lightcurve was generated using {\sc evselect} and used to examine for background flaring. The same flare seen in the RGS data was present in the PN data, thus {\sc tabgtigen} was used to filter all times when the \cps\ were greater than or equal to 10. The spectra was also examined for pileup and was found to be satisfactory. The source spectra was extracted from a circular region 35"  centred on \mrk. A background region was extracted from a circular region with a radius of 50" on the same chip. {\sc rmfgen} and {\sc arfgen} were used to generate response files for the observation. 

\subsection{\chandra}
\mrk\ was observed with \chandra\ in April 2021 for 2.1 \ks, and the data entered the archive in April 2022. The image was captured using  \chandra\ ACIS-S instrument and the data was processed using {\sc ciao 4.14}\footnote{\url{https://cxc.cfa.harvard.edu/ciao/}}. Initially a spectrum was extracted using {\sc dmextract}. The spectrum was then loaded into {\sc Sherpa} where a rough fit was obtained and saved over the energy region of interest, 0.2-2.0 \kev. The spectral shape obtained by this rough fit was imputed in {\sc ChaRT }\footnote{\url{https://cxc.cfa.harvard.edu/ciao/PSFs/chart2/runchart.html}} along with the nominal RA and DEC of the observation to generate a simulated PSF. The PSF was then projected in the detector plane using {\sc MARX} via {\sc simulate\_psf}  with a bin size of 1 pixel and a blur factor of 0.25. Next the image was exposure corrected using {\sc fluximage}, where an exposure corrected map centred around the brightest pixel in the image with a 600 pixel box, with bin size of 1 pixel. The image was extracted with a minimum energy of 0.2 \kev, a maximum energy of 2 \kev, and an effective energy of 1 \kev. Finally the image was deconvolved with the PSF generated  previously using {\sc arestore} with  the Lucy-Richardson deconvolution algorithm and 100 iterations. The finial image  presented is smoothed using a Gaussian interpolation method.
\section{Analysis}
\label{sec:results}
\subsection{Extended emission in 0.2-2 keV}
\begin{figure}
	\centering
	\includegraphics[width=\linewidth]{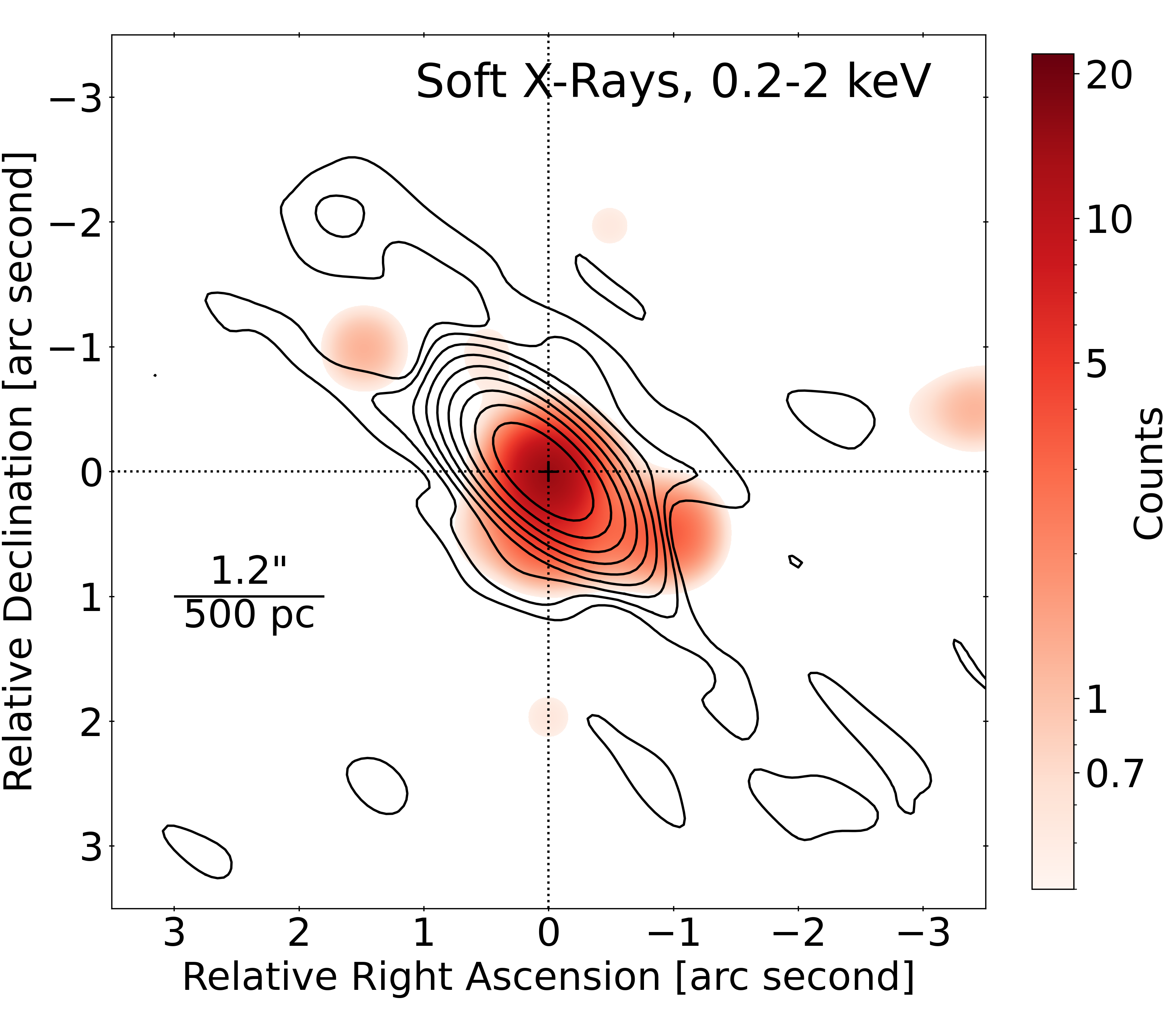}
	\caption{Snap-shot image of \mrk\ taken with \chandra-ACIS-S instrument.The JVLA 5.2 GHz radio contours are provided by  \protect\cite{Jarvela+2022}.   }
	\label{fig:june7_softChandra}
\end{figure}

The 2.1\ks\ image of \mrk\ taken with \chandra\ ACIS-S  is presented in Figure \ref{fig:june7_softChandra}. The radio contours are taken from \cite{Jarvela+2022} and are based on JVLA observations centred at 5.2 GHz with a 2 GHz bandwidth.  They  give a  sense of size and shape of the extended radio emission region.

Most of the soft X-ray emission originates from the very central region of the galaxy, suggesting at least part of it is originating from the AGN. Previous studies have indicated that there is star formation present within the central 400 pc of nucleus of \mrk\ \citep{RuschelDutra+2017}. This appears comparable to the scale of the extended soft X-ray emission in the \chandra\ image  thus some of the soft emission seen in the central region may be the result of this star formation. 
There appears to be asymmetry extending to the Southeast quadrant of \mrk.  Detailed analysis is limited by the short exposure time, and deeper observations would be needed to determine the exact location of the soft emission.   
\subsection{Emission Line Search}
\begin{figure*}
	\centering
	\includegraphics[width=\linewidth]{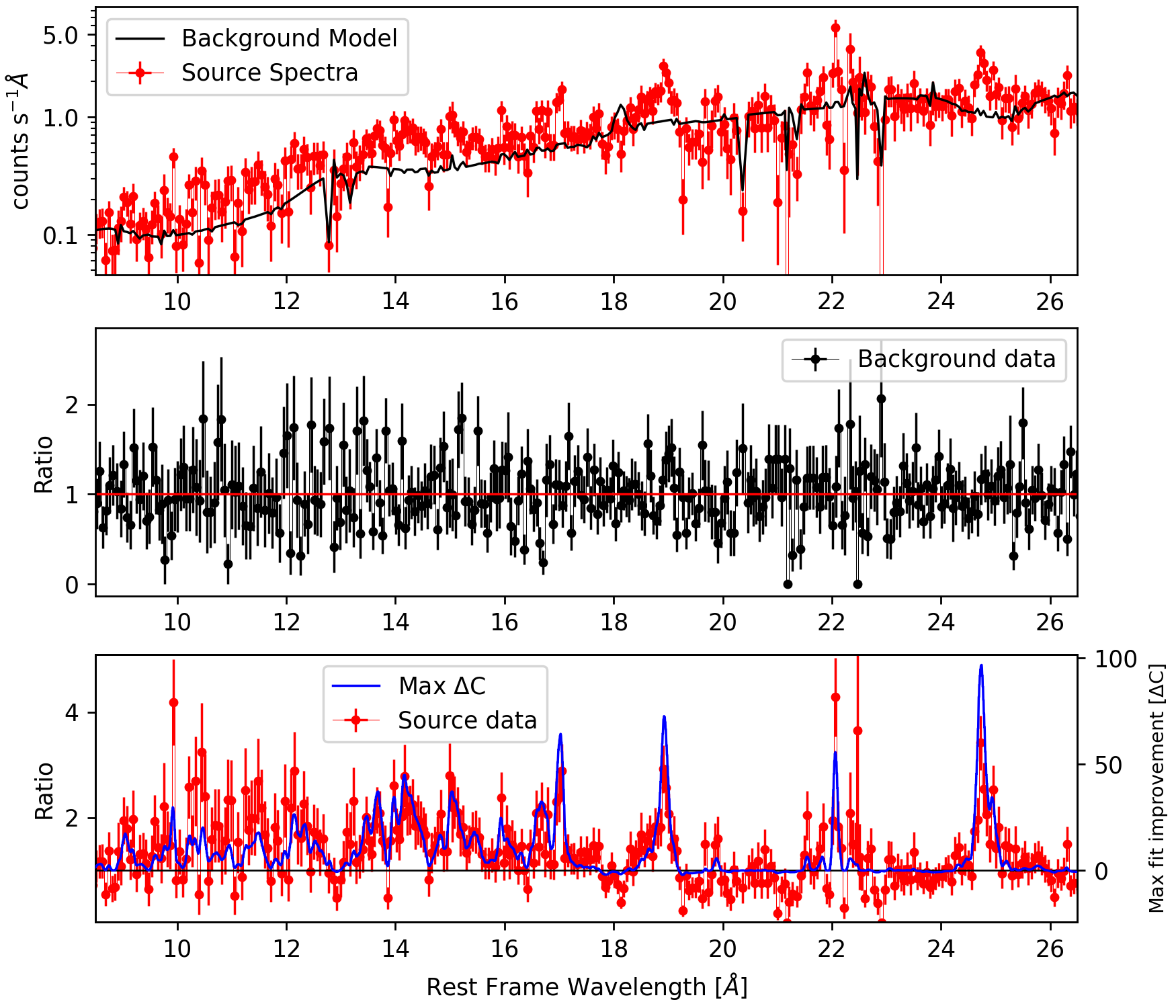}
	\caption{\textit{Top panel:} the RGS spectrum is shown in red, while the background model is shown in black. \textit{Middle panel:} the ratio of the background model to the background data. \textit{Bottom panel:} The ratio of the source data to the background model is shown in red. In blue is the maximum fit improvement (Max $\dec$) found by the line search (right axis). The black line is at a ratio of 1  on the left axis and at a maximum fit improvement of 0 C-stat on the right axis. The blue line details where the source is above the background.}
	\label{fig:aug11_dataV1_mnras}
\end{figure*}
\mrk\ is not a particularly bright source, as such we cannot confidently identify the continuum level. Instead,  we observe  emission lines poking above the background (Figure \ref{fig:aug11_dataV1_mnras} top panel). Due to the high background level it was imperative to find all regions where the data were significant over the background. This was achieved by modelling the background (see Figure \ref{fig:aug11_dataV1_mnras} middle panel) and applying the model to the source spectrum. 

A Gaussian profile with a width of $\sigma=0.01$\,\AA\ was then applied beginning at 7\,\AA, The normalization was stepped through twenty values space evenly in log space between $5 \times 10^{-7}$ and $5 \times 10^{-5}$ \anorm\ using the {\sc xspec} command {\sc steppar}. For each normalization step the best fit improvement was recorded. When complete the Gaussian profile was moved 0.005\,\AA\ (an order of magnitude finer than  the instrument resolution of RGS ) and the process was repeated, covering all wavelengths between 7 and 35\,\AA.  This allowed for the spectra to be oversampled and provide a clear picture of where the source was detected above the background. 

The maximum fit improvement was taken at each wavelength regardless of the normalisation required.  The maximum fit improvement (Max $\dec$) verse wavelength was plotted as a visualization tool to more clearly identify the locations of possible lines. This max fit improvement line can be seen in the bottom panel of Figure \ref{fig:aug11_dataV1_mnras}.  Forty-five line candidates were identified, all having a fit improvement of at least $\dec=5$. 

Next a Gaussian profile  with a width of $0.01$\,\AA\ was applied for each of the 45 line candidates, the normalizations were fit and subsequently frozen. The error on the line energies was then calculated using the {\sc xspec} error command. The uncertainty in the line energy, plus an outflow correction factor (see Sec. \ref{sec:out}), was used as the search range up until a maximum search range of 0.1\,\AA. With the line wavelength and search range  imputed in AtomDB, the most probable line candidate was identified. 

\begin{figure*}
	\centering
	\includegraphics[width=\linewidth]{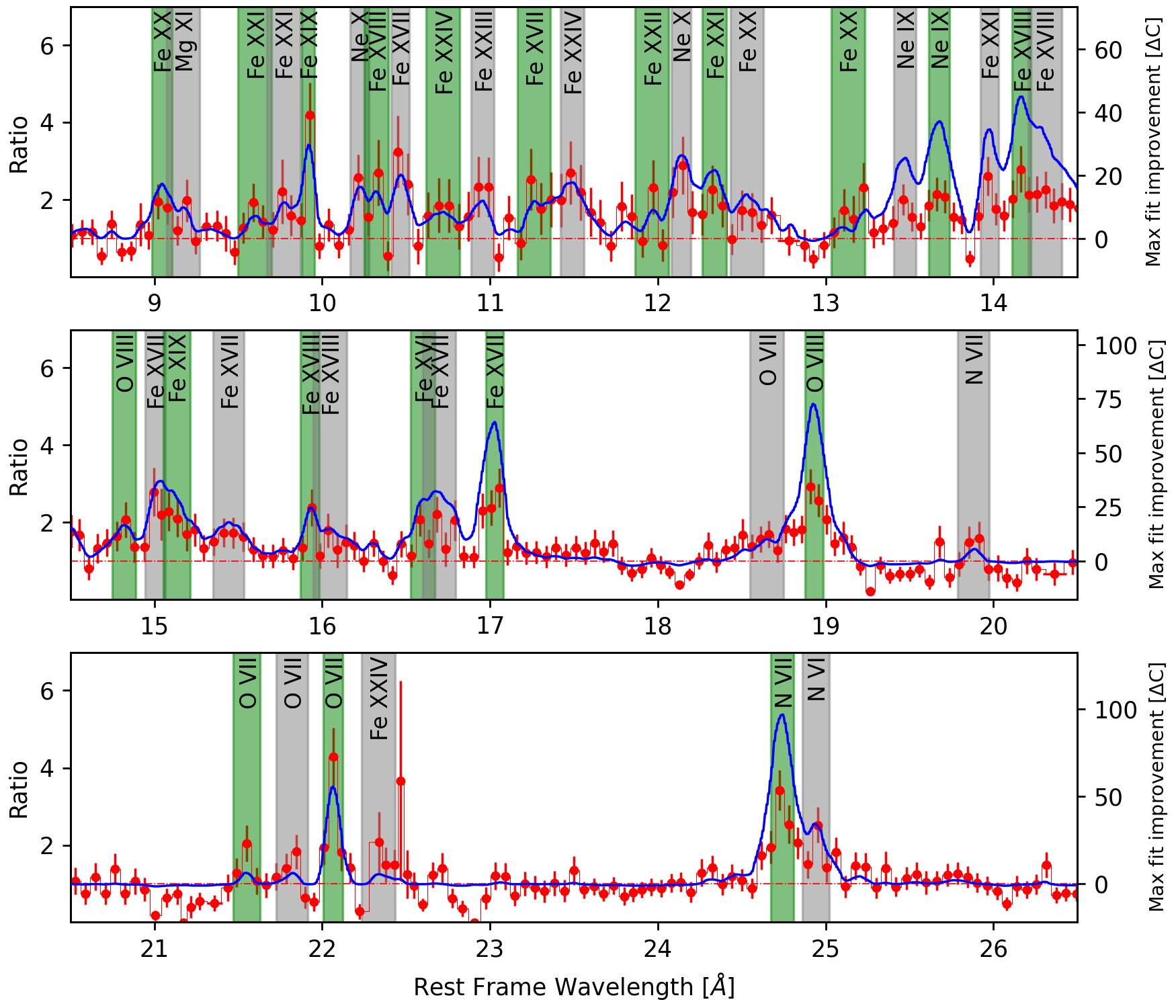}
	\caption{Results of the blind line search with candidate lines labelled. Red denotes the ratio of the RGS data to the background model. The blue line is the maximum fit improvement (Max $\dec$) at any wavelength. The shaded green and gray regions denote the search radius used in AtomDB and the label given in each shaded region is the most probable line ID that produced that emission feature. The different colour bands are simply used to distinguish nearby lines. As a reminder, anywhere the RGS data drops below 1 is a region where the background was above the source.}
	\label{fig:april18figure1}
\end{figure*}

Forty line candidates were recovered, they are detailed in Table \ref{tab:BLS}, while the data and the search line are shown in Figure \ref{fig:april18figure1}. No radiative recombination continuum (RRC) features were identified through this process as a continuum was not confidently identified. The search was preformed over the wavelength range $7-35$\,\AA, however lines were only significantly detected between $9-26$\,\AA. 
\renewcommand{\arraystretch}{1.25}
\begin{table}
	\centering
	\begin{tabular}{r c c c c c }
\hline
\multicolumn{1}{c}{(1)} 			& (2) 						& (3) 	& (4)       &(5)       	\\
\multicolumn{1}{c}{ }				& Observed					&Search &  Rest     &	       	\\
\multicolumn{1}{c}{Ion ID}	  		& wavelength	  			& range	&wavelength &$\Delta$C 	\\
									&[\AA]		  				&[\AA]  &[\AA]		& 			\\
\hline
		Fe\,{\sc xx} 	 			& $9.05 \pm 0.04$			& 0.059	&	9.066	&	17	  	\\
		Mg\,{\sc xi} 	 			& $9.2\pm 0.2 $  	 		& 0.100	&	9.169	&	10		\\
		Fe\,{\sc xxi} 	  			& $9.60\pm 0.08$	      	& 0.100	&	9.504	&	7		\\
		Fe\,{\sc xxi} 				& $9.8\pm 0.2 $		  		& 0.100	&	9.819	&	11		\\
		Fe\,{\sc xix} 	 			& $9.92^{+0.01}_{-0.02}$ 	& 0.038	&	9.945	&	30		\\
		Ne\,{\sc x} 	 			& $10.22 \pm 0.04$			& 0.057	&	10.238	&	16		\\
		Fe\,{\sc xviii}$^{\star}$	& $10.32^{+0.06}_{-0.05}$ 	& 0.072	&	10.360	&	15		\\
		Fe\,{\sc xvii} 	  			& $10.47^{+0.03}_{-0.04}$ 	& 0.053	&	10.504	&	20		\\
		Fe\,{\sc xxiv} 	  			& $10.7^{+0.1}_{-0.2}$		& 0.100	&	10.663	&	8		\\
		Fe\,{\sc xxiii}	  			& $10.96^{+0.06}_{-0.04}$ 	& 0.069	&	10.981	&	12		\\
		Fe\,{\sc xvii} 	 			& $11.26^{+0.04}_{-0.13}$ 	& 0.100	&	11.254	&	12		\\
		Fe\,{\sc xxiv} 	 			& $11.49\pm 0.05$			& 0.071	&	11.432	&	18		\\
		Fe\,{\sc xxii} 	 			& $12.0\pm 0.2$		  		& 0.100	&	11.977	&	9		\\
		Ne\,{\sc x} ($\alpha$)  	& $12.14^{+0.03}_{-0.04}$ 	& 0.059	&	12.132	&	26		\\
		Fe\,{\sc xxi} 				& $12.34^{+0.06}_{-0.04}$ 	& 0.072	&	12.284	&	22		\\
		Fe\,{\sc xx} 	 			& $12.531^{+0.1}_{-0.2}$ 	& 0.100	&	12.576	&	14		\\
		Fe\,{\sc xx} 				& $13.1^{+0.6}_{-0.3}$		& 0.100	&	12.864	&	9		\\
		Ne\,{\sc ix} ($r$)  		& $13.47^{0.05}_{-0.03}$ 	& 0.067	&	13.447	&	25		\\
		Ne\,{\sc ix} ($f$)  		& $13.68^{+0.04}_{-0.03}$ 	& 0.061	&	13.699	&	37		\\
		Fe\,{\sc xxi} 	  			& $13.98^{+0.03}_{-0.02}$ 	& 0.053	&	14.008	&	34		\\
		Fe\,{\sc xviii} 			& $14.17^{+0.03}_{-0.02}$ 	& 0.055	&	14.208	&	45		\\
		Fe\,{\sc xviii}${\dag}$  	& $14.31^{+0.03}_{-0.12}$ 	& 0.100	&	14.373	&	35		\\
		O\,{\sc viii} 				& $14.82^{+0.05}_{-0.03}$ 	& 0.070	&	14.821	&	17		\\
		Fe\,{\sc xvii} 				& $15.01^{+0.02}_{-0.04}$ 	& 0.060	&	15.014	&	37		\\
		Fe\,{\sc xix} 			  	& $15.13^{+0.06}_{-0.04}$ 	& 0.081	&	15.079	&	33		\\
		Fe\,{\sc xvii}				& $15.44^{+0.04}_{-0.08}$ 	& 0.092	&	15.453	&	18		\\
		Fe\,{\sc xviii}	 			& $15.93^{+0.02}_{-0.03}$ 	& 0.057	&	15.931	&	25		\\
		Fe\,{\sc xviii}	 			& $16.05^{+0.04}_{-0.2}$ 	& 0.100	&	16.071	&	15		\\
		Fe\,{\sc xvi}				& $16.60^{+0.05}_{-0.03}$ 	& 0.072	&	16.567	&	29		\\
		Fe\,{\sc xvii}			 	& $16.70^{+0.04}_{-0.09}$ 	& 0.099	&	16.780	&	32		\\
		Fe\,{\sc xvii}				& $17.03\pm 0.02$			& 0.052	&	17.051	&	64		\\
		O\,{\sc vii} 				& $18.65^{+0.2}_{-0.03}$ 	& 0.100	&	18.627	&	5		\\
		O\,{\sc viii} ($\alpha$) 	& $18.93\pm 0.02$			& 0.054	&	18.967	&	73		\\
		N\,{\sc vii} 				& $19.88^{+0.05}_{-0.06}$ 	& 0.094	&	19.826	&	6		\\
	  	O\,{\sc vii} ($r$)			& $21.55^{+0.04}_{-0.03}$ 	& 0.079	&	21.602	&	6		\\
	  	O\,{\sc vii} ($i$)			& $21.82^{+0.06}_{-0.04}$ 	& 0.096	&	21.804	&	6		\\
	  	O\,{\sc vii} ($f$)			& $22.06^{+0.01}_{-0.02}$ 	& 0.059	&	22.098	&	56		\\
		Fe\,{\sc xxiv} 				& $22.3^{+0.1}_{-0.2}$		& 0.100	&	22.249	&	6		\\
		N\,{\sc vii} 				& $24.74^{+0.02}_{-0.01}$ 	& 0.067	&	24.779	&	97		\\
		N\,{\sc vii} 				& $24.94^{+0.02}_{-0.04}$ 	& 0.078	&	24.898	&	34		\\
\hline
	\end{tabular}
	\label{tab:BLS}
	\caption{ (1) the emission Line ID as determined by the highest probability transition within the search range given by AtomDB. (2) The wavelength detected by the Gaussian line profile, given in Angstroms. (3) The search range used to identify the lines, it was determined by the size of the uncertainty in wavelength placement plus a z =0.002 inflow/outflow allowance up to a maximum of  0.1\,\AA. (4) The rest wavelength as reported from AtomDB. (5) The fit improvement of including this line. ($\star$) This line was fit independently of all others due to its close proximity of the Ne\,{\sc x} line at 10.238\,\AA. ($\dagger$) This line was initially frozen and then fit will all other lines free}
\end{table}
\subsection{\ovii\ Outflow velocity}
\label{sec:out}
\begin{figure}
	\centering
	\includegraphics[width=1\linewidth]{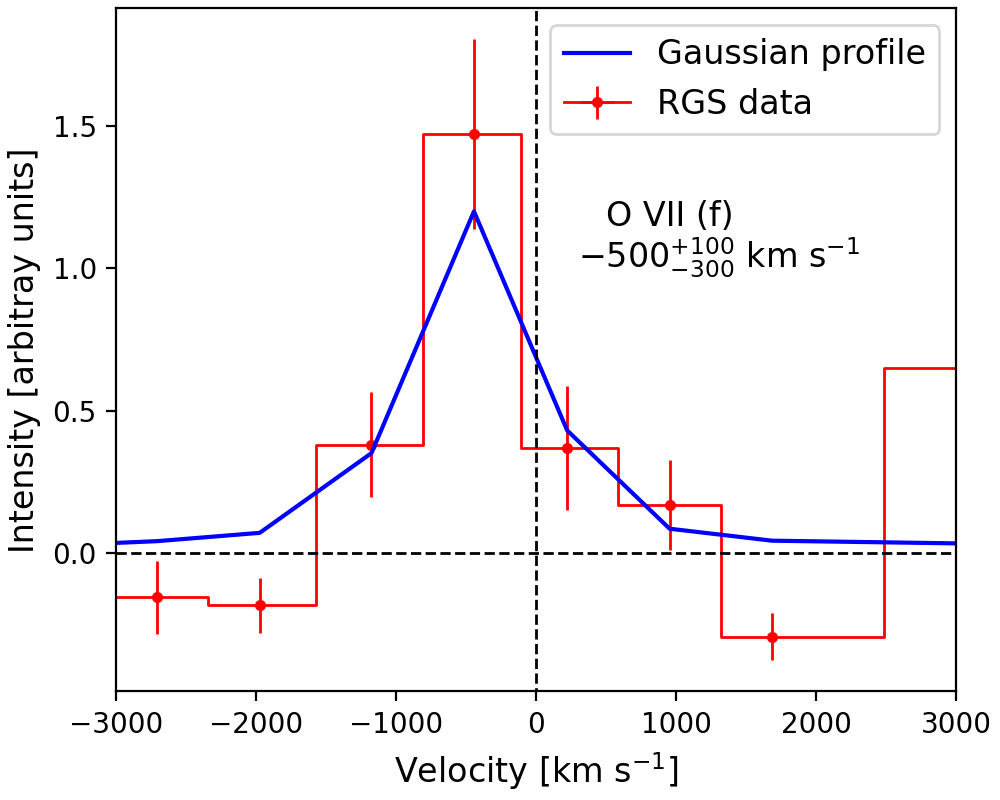}
	\caption{The velocity profile for the forbidden line of \ovii\ with a Gaussian line overlaid. The forbidden \ovii\ line is the strongest single feature in the RGS spectra, and thus is the best line candidate to explore the  outflow velocity. Based on this line the outflow velocity is  $-500^{+100}_{-300}$ \kms.  }
	\label{fig:july28oviiprofilr}
\end{figure}
While examining possible line candidates it was found that several emission features exhibited slight blue shifts in their line energy. Figure \ref{fig:july28oviiprofilr} shows the velocity profile of  the forbidden \ovii\ line. This was the strongest isolated line. The rest frame wavelength of this line is $22.06^{+0.01}_{-0.02}$\,\AA\ while the laboratory measurement is  22.098\,\AA. The resultant outflow velocity is $-500^{+100}_{-300}$ \kms. The measured velocity  is comparable to other outflows in Seyfert galaxies (see NGC\;5548 \citealt{Mao+2018}; NGC\;4151 \citealt{Armentrout+2007}; NGC\;1068 \citealt{Grafton+2021}). Outflow velocities from other emission lines were not as well constrained.  As well,  \mrk\ does show evidence for an outflow in the \oiii\ line of $\sim2000$ \kms, located between 10-100 pc of the SMBH \citep{Pan+2021}.
\subsection{Plasma Diagnostics}
He-like ions produce very specific emission features dependant on the density, temperature and processes by which the atoms were ionized. A common method for determining the density and temperature of the emitting plasma is by using the $G$ and $R$ ratios defined in Equations \ref{eq:R} and \ref{eq:G}. For these diagnostics $z$ is the  forbidden line ($f$), $x+y$ are the inter-combination lines ($i$), and $w$ is the resonance line ($r$) \citep{Porquet+2000}.
\begin{equation}
\centering
R = \frac{z}{x+y} = \frac{f}{i}
\label{eq:R}
\end{equation}
\begin{equation}
G  = \frac{x+y+z}{w} = \frac{f+i}{r}
\label{eq:G}
\end{equation}
For \mrk\ only  the He-like triplet of  \ovii\ is strongly detected. There is some evidence of Mg\,{\sc xi} and \neix, but the \ovii\  triplet is clearly seen. This may be due to the blend of lines around Mg and Ne originating from other species and the relative lack of other lines around the \ovii\  triplet. For this work both the \ovii\ and \neix\ He-like triplets are explored. 

Due to the weak intercombination lines in this spectra it was difficult to ascertain the uncertainties of the $G$ and $R$ ratios. To overcome this we implemented a model in {\sc xspec} so that the values and their uncertainties might be directly outputted. This was achieved by re-writing the $G$ and $R$ ratio such that $f$ and $r$ could be expressed in terms of $G$, $R$ and $i$. To do this the first step is a simple rearrangement of the $R$ ratio:
\begin{equation}
f = R\times i \;.
\label{Eq:f}
\end{equation}
Next the $G$ ratio: 
\begin{equation}
r  = \frac{f+i}{G} \;.
\end{equation}
Substituting in what we know about $f$ from Equation \ref{Eq:f};
\begin{equation}
r  = \frac{R\times i+i}{G}
\end{equation}
The subtlety of this method is revealed when we consider how to implement it directly in {\sc xspec}. The expression would be: $(constant_G\times zagauss_r + zagauss_i + constant_R \times zagauss_f)$, where the subscripts denote the parameter the component represents. The correct normalization for $zagauss_r$ is $\frac{R\times i+i}{G}$ but the component is multiplied by $G$ therefore the normalization must be $\frac{R\times i+i}{G^2}$. Similarly the correct normalization on f is  $R\times i$, because the component is already multiplied by $R$ we simply link the two components. This leaves $G$, $R$ and $i$ free to fit the triplet.  The added complication here is the outflow we see on \ovii($f$), thus we have an addition component, {\sc vmshift}, to account for the outflow. The outflow velocities are linked for the $r$, $i$ and  $f$ lines. The $G$ and $R$ ratio for \ovii\ and \neix\ as well as their outflow velocities are displayed in Table \ref{tab:GR}. 
\begin{table*}
	\begin{tabular}{c c c c c} 
		\hline
		(1)								  & (2)									 &     (3)		&   (4)				&	(5)					\\
		Model Component   & Model Parameters	& Short name	& \ovii\ 			& \neix\			\\
		\hline
		Resonance Line  	 & Wavelength 				   & $\lambda_r$	& 21.602\,\AA		& 13.447\,\AA	\\
											 & Normalization	 	& $r$			& $=(i+i*R)/G^2$	& $=(i+i*R)/G^2$ \\
		\hline
		Intercombination Line	& Wavelength 		& $\lambda_i$	& 21.804\,\AA 		& 13.553\,\AA	 \\
											& Normalization	 	& $i$			& $10^{-5}$			& $3\times10^{-6}$\\
		\hline
		Forbidden Line		 & Wavelength  		& $\lambda_f$	& 22.098\,\AA		& 13.699\,\AA 	 \\
											& Normalization	 	& $f$			& $=i$				&$=i$					\\
		\hline
		G 								  &			 		& $G$			&$5^{+15}_{-3}$		&$1.5^{+3.5}_{-0.9}$			\\
		{\sc constant} 		  &					&				&					&						\\		
		\hline
		R 								  & 					& 	$R$			&	$>2.5$			&	$>1$				\\
		{\sc constant} 																							\\	
		\hline	
		Outflow Velocity    & 					& 				&	$-500\pm200$	\kms	&	$-300^{+800}_{-600}$ \kms				\\
		{\sc vmshift} 																							\\	
		\hline	
	\end{tabular}
\label{tab:GR}
\caption{G and R diagnostic ratios for \ovii\ and \neix. The model components are listed in column (1), the model parameters are listed in column (2) and their short form names used in the text are listed in column (3). The values for each parameter for the \ovii\ and \neix\ triplets are given in columns (4) and (5), respectively.  }
\end{table*}

Comparing the strength of the resonance line to the strength of the Ly${\alpha}$ line reveals that the ratio of H-like/He-like atoms for both the \ovii\ and \neix\ atoms is of order unity. This allows us to use the G ratio to find a temperature of the order $10^6$ K \citep{Porquet+2000}. This estimation is not with out its caveats, as it assumes that the regions where the H-like ions are formed is the same as where the He-like ions are formed which is not necessarily the case according to \cite{Porquet+2000}. 

Due to the weak intercombination line in both triplets the R ratio was not well constrained and we can only provide lower limits on it and the density estimations. Comparing the calculated values to the $R(n_e)$ curves produced by \cite{Porquet+2000} we gain limited information. For \ovii, which has $R>2.5$, this tells us that the plasma that produced these emission lines has a density less than $10^{11}$\cm$^{-3}$. The plasma that produced the \neix\ emission lines has a density less than  $10^{12}$\cm$^{-3}$.  Moreover, the outflow velocity of the \ovii\ triplet is consistent with the velocities reported in Section \ref{sec:out}. The outflow velocity of the \neix\ triplet  agrees with that of the  \ovii\ triplet, and it also agrees with being at rest with the galaxy as a whole. 

The finial piece of information we can deduce from the He-like ions is from the relative strength of the resonance line.  In purely photoionized plasmas the resonance line is weaker compared to the intercombination and forbidden lines. Whereas, in mixed plasmas, containing both photoionized and collisionally ionized emission, the resonance line is relatively strong. If we look at the \ovii\ triplet we see an incredibly strong forbidden line and weak intercombination and resonance lines. Assuming we are in the region where $R$ is very sensitive to density (which was allowed by the calculated $R$ ratio) than this is indicative to a photoionized plasma. 

\subsection{Neon Abundance}
\label{sec:noNeon} 
A neon overabundance was proposed in \cite{Grupe+2004} to explain a feature they observed at 0.91 $\kev$ in the 5\ks\ PN data. The feature was attributed to the \neix\ triplet. They did not detect the \ovii\ triplet located at $\sim$ 0.56 \kev\ leading them to conclude that there was an overabundance of Ne in \mrk. Subsequent works have used this overabundance while fitting the X-ray spectra (see \citealt{Jiang+2021}). 

\begin{figure}
	\centering
	\includegraphics[width=\linewidth]{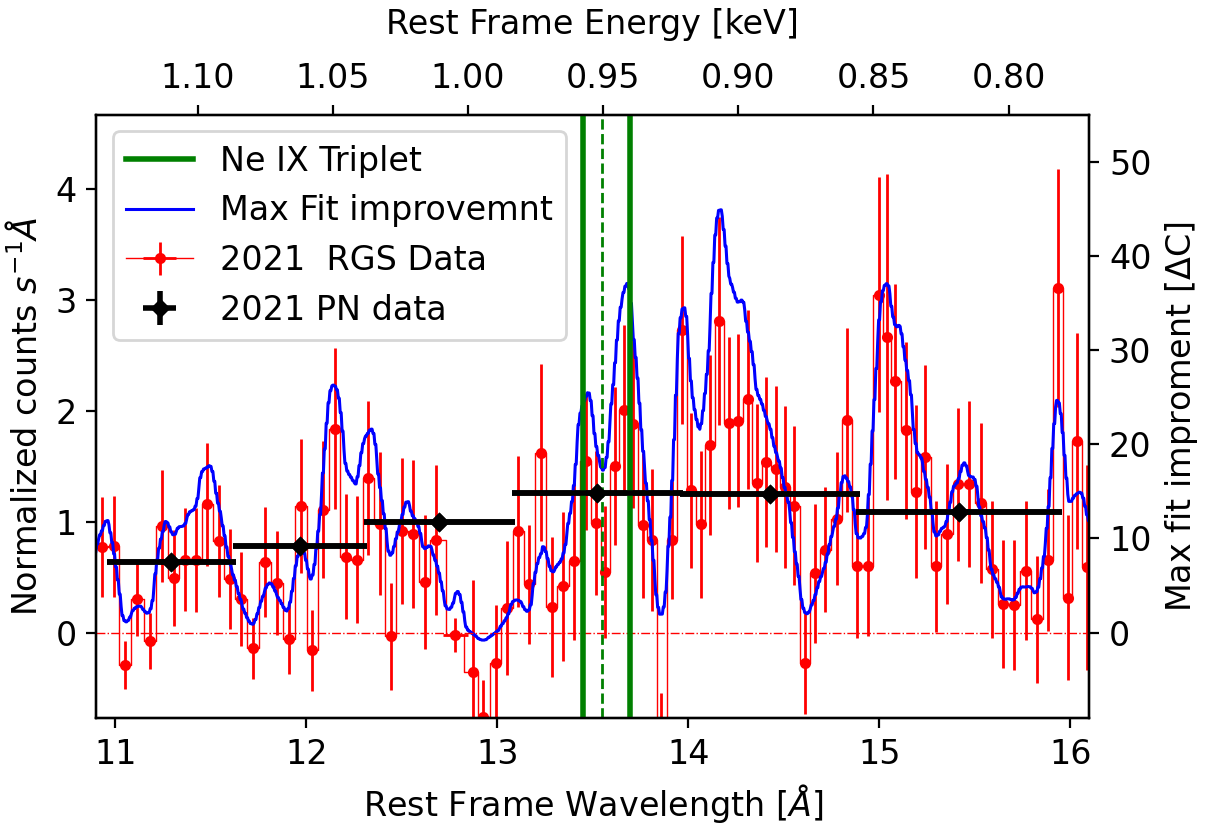}
	\caption{Comparison of the spectral resolution of the EPIC PN instrument to the RGS instrument. Red shows the RGS data, blue is the maximum fit improvement line (same as Figure \ref{fig:aug11_dataV1_mnras}) black denotes the PN data and green shows the location of the \neix\ triplet. The PN and RGS spectra have been normalized so that they may be presented on the same scale.}
	\label{fig:Neon}
\end{figure}

The analysis of \cite{Buhariwalla+2020} suggested the excess emission at 0.91 \kev\ was in fact due to a blend of emission lines including the \neix\ triplet and Fe-L emission lines. In Figure \ref{fig:Neon}  a comparison between the RGS spectra and the EPIC PN spectra is made. Here we see both the RGS (in red) and PN spectra (in black) compared from $11-16$\,\AA. Both spectra have been normalized by their mean in this region to ease the comparison.

The red RGS data points show a plethora of emission features while the black PN data show an excess emission between 0.85 and 1 \kev, similar to what was seen in the original 2001 observation by \cite{Grupe+2004}. Even with the deep 2021 PN observation, the \neix\ emission lines remain unresolved. The RGS spectrum is required to resolve the \neix\ triplet and the surrounding features to  state that the excess emission seen in this object at 0.91 \kev\ is not due solely to the \neix\ triplet but instead a it is a blend of many emission features in this region. 

Furthermore the overabundance described by \cite{Grupe+2004} was reliant on the non-detection of the \ovii\ triplet in addition to the enhancement of the \neix\ line by Fe-L emission. In the bottom panel of Figure \ref{fig:april18figure1} we can clearly see the \ovii\ present in our RGS spectra. The detection of the \ovii\ triplet and the confirmation that the \neix\ triplet was being enhanced by surrounding emission lines the shows a Ne overabundance is not needed in this source. 
\section{Spectral Modelling}
\label{sec:spectral}
The RGS data were background modelled in {\sc xspec}, and C-statistics were used \citep{Cash+1979} to evaluate the fit quality throughout. Errors were calculated at the 90 \% confidence level using the {\sc xspec} error command. Based on the results of the plasma diagnostics we proceeded with the assumption that there were at least two ionized regions present in the spectra. 

We tested collisionally ionized plasma with  \apec, version 3.0.9, \citep{Smith+2001} and a photoionized plasma using  {\sc xstar} grids, version 2.31 \citep{Kallman+2001}. The model  \mekal\ \citep{Liedahl+1995} was tested in this analysis as it was what was originally used to model the data in \cite{Buhariwalla+2020}. Ultimately \apec\ was used as it contains more up to date atomic data, and produced better fit statistics. All fits were absorbed  with a Galactic  column density of \nh$=4.43\times10^{20}\; \text{atoms/cm}^2$ \citep{Willingale+2013}.

{\sc xstar} grids were calculated with a covering fraction of 1, luminosity of $10^{42}$\ergps\ and the turbulent velocity was fixed at 300 \kms. The ionizing continuum was described by a powerlaw with $\Gamma=2.2$.  The column density was logarithmically sampled at five points between $10^{20}$ and $10^{24}$ cm$^{-2}$. The density of the emitting material was logarithmically sampled at six points between $10^9$ and $10^{14}$ cm$^{-3}$  The ionization parameter is defined as $\xi = L/nr^2$ \citep{Tarter+1969, Kallman+2001} where $L$ is the ionizing luminosity, $n$ is the density of the material and $r$ is the distance between the ionizing source and the emitting material. The log ionization was sampled linearly at 6 points between $\text{log} \xi$ 1 and 4 erg cm s$^{-1}$. The column density, ionization and density were all interpolated for a total of a 300 step grid.

Seven subsequent grids were created with density at a fixed value ranging from $10^4$ to $10^{10}$ cm$^{-3}$, \, log$\xi$ was linearly samples at 10 points between 0 and 5, the column density was logarithmically sampled at ten points between $10^{20}$ and $10^{24}$ cm$^{-2}$. All other parameters were kept the same. These were 100 step grids. The $10^9$ and $10^{10}$ cm$^{-3}$ grids were tested against the 300 step grid for consistency, they produced the same fit statistic and ionization parameters. All abundances were fixed at solar.

For  \apec\  the plasma temperature ($kT$) and normalization remained free, while the abundances remained fixed at solar. The powerlaw, when included, had free photon index and normalization.

Initially, emission from only one plasma component (\apec\ or  \xstar)  was fitted to the RGS spectrum between $7-31$\,\AA. With each individual plasma, the addition of a powerlaw component was considered to account for the underlying continuum. 
Fitting each of these components individually was able to account for select emission lines, but not all the emission features. The fits were $\cdof=1319/814$ for \apec\ and $\cdof=1360/813$ for \xstar. The addition of a powerlaw ({\sc po}) component had fit improvements of   $\dec=22$ for  \apec+{\sc po} for two additional free parameters. The addition of the powerlaw to \xstar\ had a drastic fit improvement of $\dec=83$ for two additional free parameters, but this was due to the poor initial fit of the \xstar\ model \xstar+{\sc po} results in a fit of $\cdof=1277/811$.

Next two component plasmas were tested where both components were of the same type, this was done to explore the possibility of a single plasma with a gradient of temperature and/or ionization values. This produced slightly better fits than the addition of the powerlaw, with the  fit improvement for \apec+\apec\  was $\dec=27$ the single component counterpart, for two additional free parameters. \xstar+\xstar\ resulted in a fit improvement of $\dec=69$, producing a worse fit than a single component+powerlaw model. \cite{Jiang+2021} fit the the soft band of \mrk\ with two collisionally ionized components. However their CCD spectra lacked the data quality to  distinguish between a collisionally ionized and photoionized component. 

\subsection{Testing for Two Distinct Plasmas }
\label{subsec:two_plasma}
\begin{figure}
	\centering
	\includegraphics[width=1\linewidth]{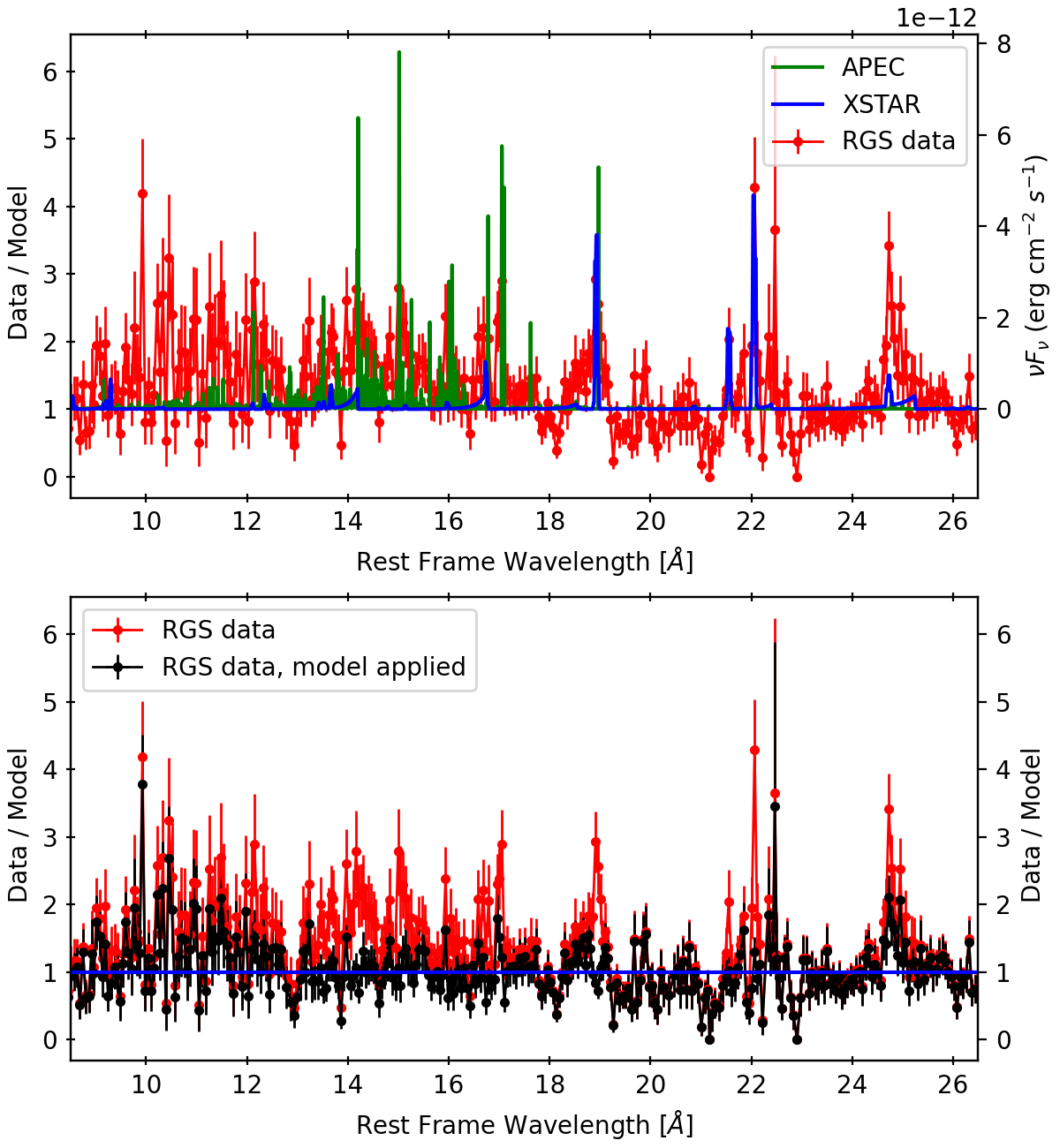}
	\caption{\textit{Top panel:} the ratio of the RGS data and background model is shown in red, the fit here is \cdof= 1716/816. All data above ratio$=1$  is above the background level. The best fit two component plasma  is shown in green ({\sc apec}) and blue ({\sc xstar}). The individual models are plotted to show which features are fit by each. \textit{Bottom panel:} the red data are the same as above. The black data are the ratio of red data with the best fit model folded through the response (\cdof=1158/810).  }
	\label{fig:aug8fig3}
\end{figure}

A mixed plasma was tested such that it contained a collisionally ionized (\apec) and a photoionized  (\xstar) component.  This produced a fit of  \cdof=1201/811. The \apec\ component produces a $\dec \sim15$ better than the \mekal\ component for the same degrees of freedom. The \ovii$(f)$ line, which was outflowing with a velocity of $\sim-500$ \kms, was produced by the photoionized plasma. Thus we test the \xstar+\apec\ models with the photoionized component allowed to outflow.  The {\sc vmshift}\,$\times$\,\xstar\,+\,\apec\ model produces a superior fit of \cdof=1156/810, $\dec=45$ for one additional free parameter. Similar fit improvements were seen in the \xstar+\xstar\ and \apec+\apec\ fits when one component is allowed to outflow. No significant fit improvement were seen in these fits when both components are allowed to outflow. The best fit model can be found in Figure \ref{fig:aug8fig3},  the best fit parameters can be found in Table \ref{tab:xstar}. 

\begin{table}
	\centering
	\label{tab:xstar}
	\begin{tabular}{c c c } 
		\hline
		(1)		&(2)								&(3)					\\
Model Component	& Model Parameter					& Value 				\\
		\hline
		\apec 	& kT [keV]										& $0.74_{-0.07}^{+0.06}$\\
		 		& Abundance 										& $1^f$					\\
		 		& Normalization  [$\times10^{-5}$]  & 		$3.8\pm 0.6$			\\
		\xstar 	& Column density [cm$^{-2}$]& $<\num{2.5e22}$	  	\\
		 		& Log Ionization [erg cm s$^{-1}$]	& $1.7_{-0.2}^{+0.3}$ 	\\
		 		& Density 	[cm$^{-3}$]						& $>\num{3.2e10}$     	\\
		 		& Redshift												& $0.01993^f$		  	\\
				& Normalization	[$\times10^{-6}$]	 & 		$2.6_{-0.4}^{+0.5}$ \\
  {\sc vmshift} & Velocity	[km/s]					& 	$-660\pm150$	  	\\
		\hline	
	\end{tabular}
	\caption{Best fit model parameters for the two component plasma model. Column (1) indicates the model component,  Column (2) gives the Model Parameter and  Column (3) gives the value of each parameter. All parameters with the superscript `$f$' are kept fixed at quoted values. }
\end{table}
We can see that the photoionized material is outflowing with a velocity of $-660\pm150$ \kms\ which is consistent with the outflow velocity that we saw with \ovii($f$) emission line. The density of the emitting material is $>\num{3.2e10}$ cm$^{-3}$, which is consistent with the density estimation found using the \ovii\ $R$ ratio. The temperature of the emitting plasma is consistent with the temperature found in \cite{Buhariwalla+2020}, despite the difference in model component (\apec\ vs \mekal) and the difference in instrument (RGS vs MOS). There appears to be an excess of emission below 12\,\AA\ and the $\sim25$\,\AA\ feature attributed to N{\sc vii} appears under fit.

\begin{figure}
	\centering
	\includegraphics[width=1\linewidth]{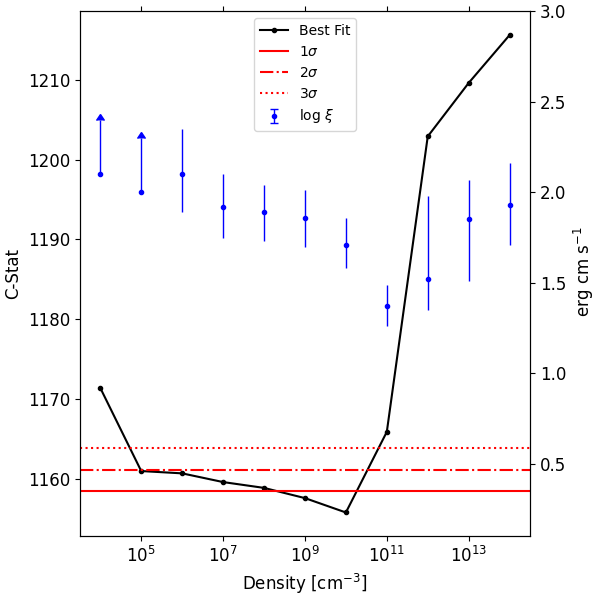}
	\caption{ Density as a function of C-Stat (left axis) for the best fit two component plasma at different densities, and measured ionization at different density values (right axis).}
	\label{fig:sept29_Figure_2}
\end{figure}

The best fit two component plasma gave an upper limit on density of less than $\num{3.2e10}$ cm$^{-3}$, the \xstar\ grid that gave this result was limited to densities above $10^9$ cm$^{-3}$. Thus the lower limit of density was estimated using previously made grids with the same parameters.   Figure \ref{fig:sept29_Figure_2} (left axis) shows the C-stat as a function of density, the minimum C-stat is 1156, given by the best fit model. 1, 2, and 3 $\sigma$ lines are drawn to show the allowed density of the photoionized plasma.  The allowed density is then between $10^5 - 10^{10}$ cm$^{-3}$. Ionization as a function of density (Figure \ref{fig:sept29_Figure_2} right axis) will be discussed in section \ref{subsec:PIP}.

{\sc xtar}  grids with ionizing continuum described by a powerlaw with  $\Gamma=1.7$ and $\Gamma=2.5$ were generated to test the effects of  the continuum shape on the spectral appearance. They produced similar fit statistics as the best fit model, with $\Gamma=2.2$. The measured properties of the {\sc xtar}  and {\sc apec} components  were consistent, within error, to those quoted in Table \ref{tab:xstar}.  This led us to conclude that the shape of the continuum  has a minimal effect on the properties of the photoionized emitter. This does not seem unreasonable in \mrk\ due to the modest data quality and small number of emission lines fit by the {\sc xstar } component.

\xstar\ grids with variable abundances were generated to explore the possibility of an overabundance in \mrk. To ease computational time the density was fixed at $10^6$ cm$^{-3}$ and column density was fixed at $10^{23}$ cm$^{-2}$. All other parameters remained the same. Grids with variable Ne, O, Fe, and N were generated. Variable Ne and Fe  had no affect on the fit of the RGS spectrum.  Variable O and N produced best fit models with super solar ratios of \NOx. The best fit grid produced a relative abundance of $\NOx =4\pm2$ $A_{\odot}$, with a $\dec=16$. After investigation it was found that the driving mechanism for this super solar abundance is the 25\,\AA\ N{\sc vii} feature that was under fit in Figure \ref{fig:aug8fig3}. When the variable N grids are fit to the data below 25\,\AA\ no overabundance or fit improvement is found.   However  when the data above 25\,\AA\ was included the overabundance is present. \cite{Mao+2019} showed that the relative \NOx\ abundance could be super-solar in the ICM due to enrichment from low and intermediate mass stars. Their measured \NOx\ agree with the value found here. However it is not immediately clear if this enrichment effect  could also be present in the photoionized material of  \mrk, or if some other processes is driving the excess emission at 25\,\AA.

The model under fits the data below 12\,\AA,  and no combination of collisionally or photoionized plasma could sufficiently fit the data in this region.  This excess emission is also seen in the PN spectrum from this observation (Buhariwalla \et\ in prep.) indicating that the excess emission is probably not from statistical fluctuations. Instead it may be that the AGN continuum in  \mrk\ is becoming visible at wavelengths less then 12\,\AA\ (energies greater than 1\;\kev). The RGS data are insufficient to make any inferences about the continuum. Thus this feature will be followed up in the future work on \mrk\ and its broadband spectra. 
\section{Discussion}
\label{sec:disc}
\subsection{The Origin of the Photoionized Plasma}
\label{subsec:PIP}
\begin{figure}
	\centering
	\includegraphics[width=1\linewidth]{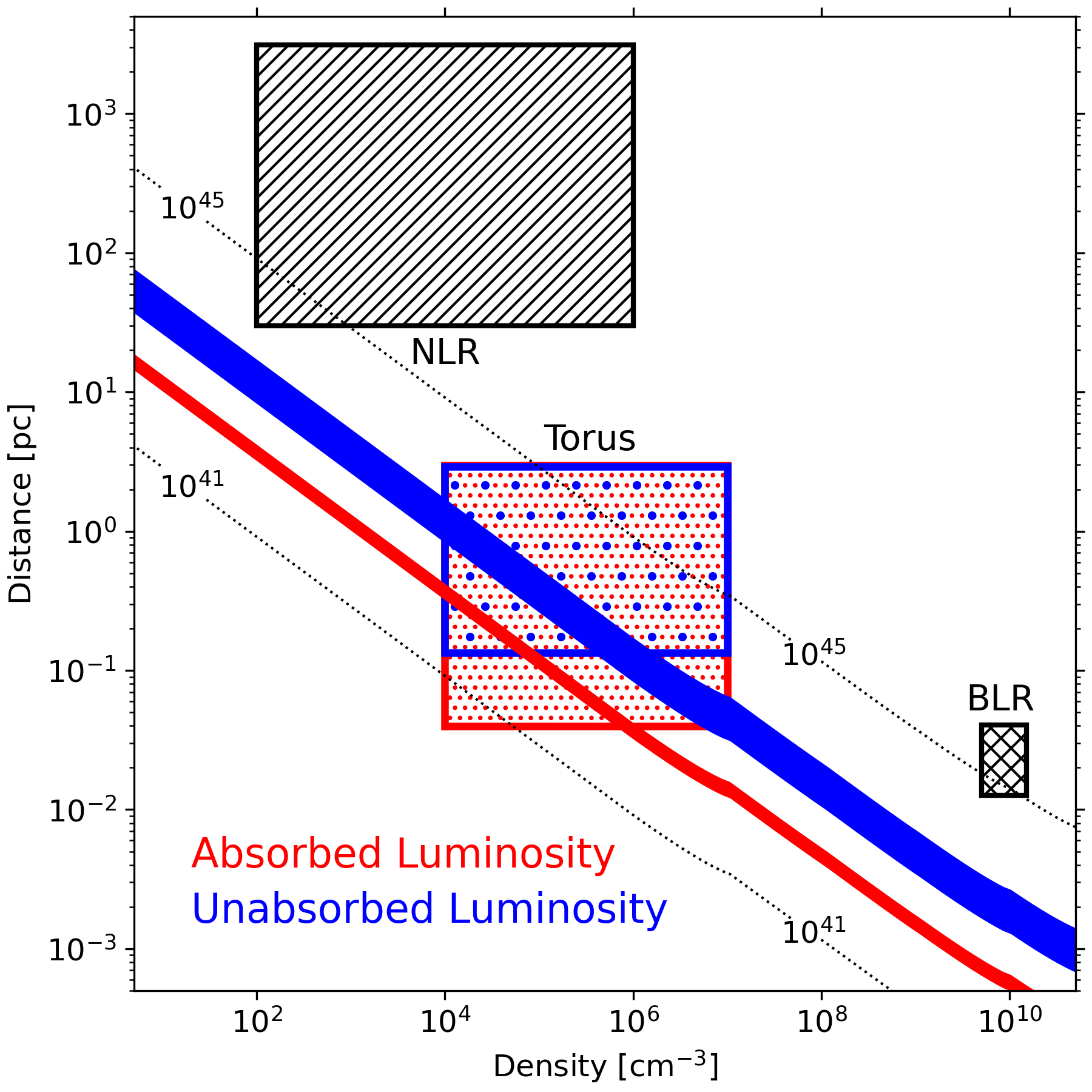}
	\caption{Density as a function of distance outlying the density-distance combinations that would produce the correct ionization for a given luminosity. The boxes outline the density-distance combinations for known components in  AGN,  they include the narrow line region (NLR), the broad line region (BLR) and the torus. See Section \ref{subsec:PIP} for further details.  }
	\label{fig:sept29_Figure_1}
\end{figure}
To explore the origin of the photoionized emitter we examined properties that could result in the measured ionization parameter. Recall the  definition of the ionization parameter $\xi = L/nr^2$ (see Sec. \ref{sec:spectral}). This  can be rearranged as $r(n) = \sqrt{L/n \xi}$, allowing us to plot distance as a function of density. The ionization parameter ($\xi$) is degenerate with the density of the emitting material. This degeneracy is mitigated with the use of multiple \xstar\ grids generated at different plasma densities. The right panel (right axis) of Figure \ref{fig:sept29_Figure_1} shows $\text{log} \xi$ as a function of density.

The ionizing luminosity ($L$) was estimated using the models in \cite{Buhariwalla+2020}, however due to the uncertainty in the location of the emission material it is undetermined if the photoionized emitter sees the absorbed or unabsorbed luminosity, thus both luminosities are tested. Although the shape of the ionizing continuum would change between the absorbed and unabsorbed scenarios we showed in Section \ref{sec:spectral} where dramatic changes in the continuum have a minimal effect on the properties of the photoionized emitter. Thus for this exercise both luminosities can be used to estimate the properties of the photoionized emitter. We caution that the solution for the unabsorbed luminosity is to be taken as  approximate, as we did not re-run the models using the absorbed continuum to illuminate the emitting gas at all densities. A single \xstar\ grid with density $10^{10}$ cm$^{-3}$ was constructed using an absorbed ionizing continuum based on the absorbed continuum from \cite{Buhariwalla+2020}. It produced an ionization parameter greater than the unabsorbed scenario, however the difference was small on the distance-density plot. It resulted in a slight shift downward of the allowed regions for the absorbed luminosity.
 
The blue band in Figure \ref{fig:sept29_Figure_1} (left panel) shows where allowed combinations of distance and density lie for an unabsorbed source. The  red band shows the same for an absorbed source. The variable $\xi$ has some influence on the $r(n)$ plot, causing a small wave-like effect on the bands in Figure \ref{fig:sept29_Figure_1}.  To illustrate what kind of material would be able to produce these emission lines density and distance combinations of known AGN components are drawn in this plane.  

The outer radius of the NLR was estimated using   a $\textrm{R}_{\textrm{NLR}} - \textrm{L}_{\textrm{\oiii}}$ relationship derived for a sample of quasars and Seyfert galaxies \citep{Bennert+2002}.  The value of $\sim3$ kpc is consistent with the radius of the extended NLR in \mrk\ found by \cite{Husemann+2022}. We take the inner radius of the NLR to be 30 pc, obtained from multi-density photoionization modelling of the NLR of Seyfert galaxies that reproduces a large range of NLR emission line intensities \citep{Komossa+1997}. This is consistent with a sample of Seyfert 2s and intermediate Seyfert ($\sim$\ Sy1.5) galaxies \citep{Vaona+2012}, and estimates using a second $\textrm{R}_{\textrm{NLR}} - \textrm{L}_{\textrm{\oiii}}$ relationship for the effective radius of the NLR in Seyfert galaxies given by \cite{Schmitt+2003}. For \mrk, comparable values of L$_{\textrm{\oiii}}$ are  given by \cite{Rafanelli+1984} and \cite{Malkan+2017}.  The density is estimated measurements of intermediate Seyfert, and  Seyfert 2 galaxies \citep{Bennert+2006, Vaona+2012}.  We note that the density and inner radius of the NLR may extend to lower values, but this provides a good order-of-magnitude estimate for our work. 

The BLR size was estimated using R$_{\textrm{BLR}} - \lambda \textrm{F}_{\lambda}(5100 \, \angstrom)$ \citep{Kaspi+2005}, with $\lambda \textrm{F}_{\lambda}(5100\, \angstrom)$ for \mrk\ taken from \cite{Grupe+2004} and \cite{Pan+2021}. The density of the BLR  is given by \cite{Netzer+2013, Arav+1998}.  

The outer radius of the torus in \mrk\ is measured to be 3 \pc\ based on 12 $\mu m$ observations \citep{Tristram+2011}. The inner torus radius  (0.1-0.01 pc) is estimated  by the dust sublimation radius with the luminosity estimated using models from \cite{Buhariwalla+2020}. The density of the torus is taken from \cite{Netzer+2013}.  

With all these components placed in the density - distance plane we begin to see what kind on material is producing the photoionized emission in \mrk. A tours like structure produces the correct density - distance combination in both the absorbed and unabsorbed scenario. As discussed in section \ref{subsec:two_plasma}  the allowed density is  between $10^5 - 10^{10}$ cm$^{-3}$, which agrees with a tours like structure again. The farthest allowed emitting region would be dependant on the lowest allowed density ($10^5 \; \textrm{cm}^{-3}$) and it would be at most a few pc, around the measured outer edge of the torus in \mrk\ \citep{Tristram+2011}.

If we analyses the full width half max (FWHM) of the \ovii\ ($f$) line we fine that FWHM$_{\textrm{\ovii}\;(f)}$ <520 \kms. The Keplerian velocity of an object orbiting around the SMBH at the inner radius of the torus is $\sim$ 500 \kms. We can then conclude that the photoionized emitting material cannot be located any closer than the inner radius of the torus. 

\begin{figure*}
	\centering
	\includegraphics[width=1\linewidth]{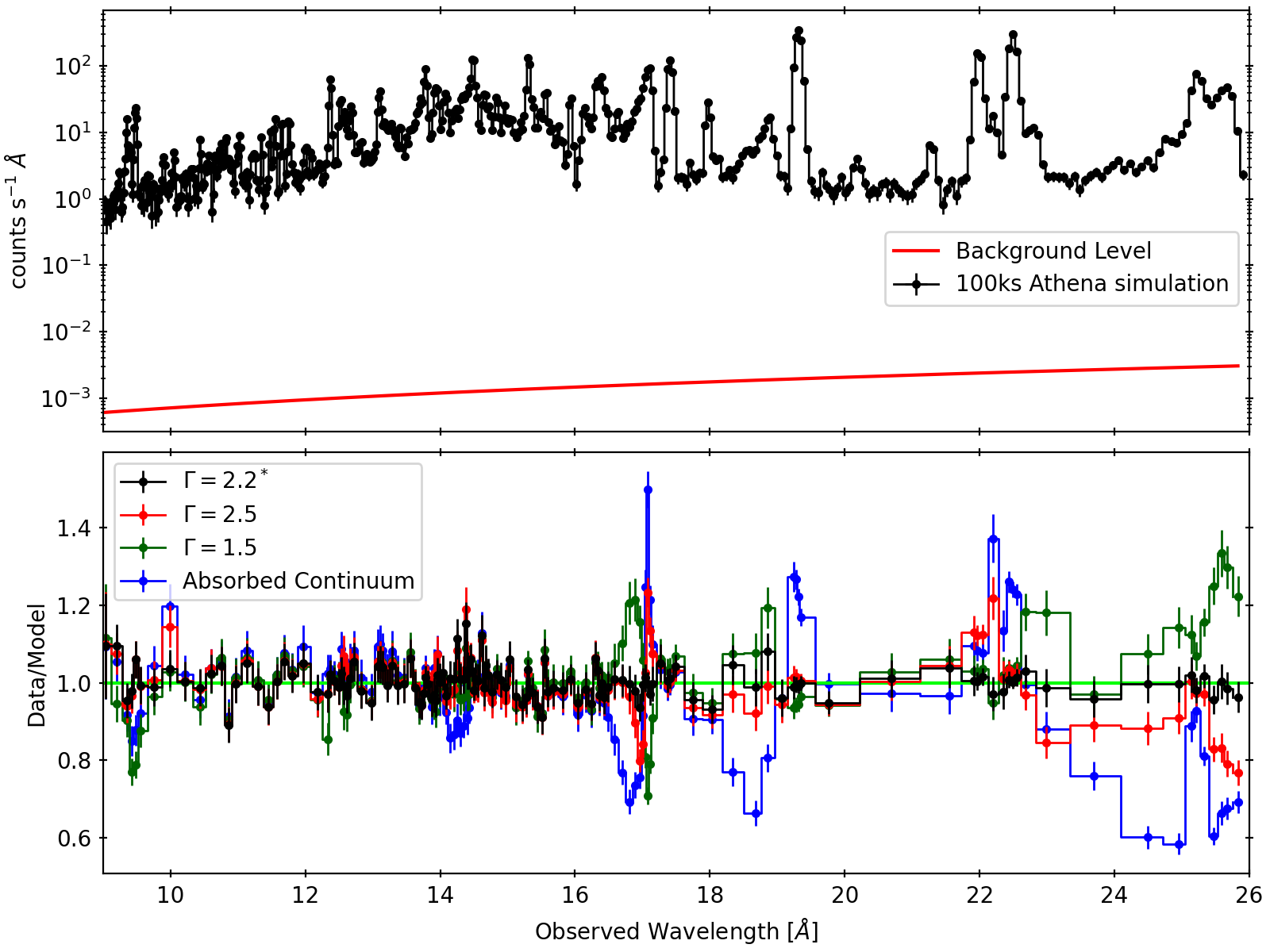}
	\caption{ \textit{Top panel:} 100\;ks \textit{Athena-XIFU} simulation with the best fit \xstar+\apec\ model. The \xstar\ grid was illuminated by a powerlaw continuum with $\Gamma=2.2$.   The approximate  \textit{XIFU} background is shown in red. \textit{Bottom panel:} the Athena simulation fit with \apec\ +\xstar, for four different \xstar\ grids. The four grids have the same input parameter (covering fraction, luminosity, turbulent velocity, density, column density, ionization and abundances) but are illuminated by different ionizing continuum. The ionizing continuum are a mathematical powerlaw  with $\Gamma=2.2$, $\Gamma=2.5$ and $\Gamma=1.5$, as well as an empirical  continuum model based on the absorbed continuum from the broad band spectra of \mrk\ \protect\citep{Buhariwalla+2020}. The ratios have been binned up for visual clarity.  }
	\label{fig:dec19_fig1}
\end{figure*}
In summary,  the RGS data have greatly increased our knowledge of the photoionized plasma in \mrk, allowing us to determine its ionization, density and location. We now know that the material exists  within several parsec of the central engine. Previous to these results it was not abundantly clear that there were emission line features in \mrk, let alone to the extent detected here. However, several questions about the physical descriptions of the  few parsec surrounding the central engine of \mrk\ remain. Namely, is this photoionized material related to the absorbing material we know is present in this source? Modelling the broadband X-ray continuum of \mrk\ requires heavy absorption from neutral and warm absorbers \citep{Buhariwalla+2020}. It is easy to image that a large cloud could be illuminated by the central engine obscuring our line of sight to the central engine, while other regions of the cloud are  photoionized, producing the photoionized spectra we see here. 

Broadband analysis will   shed some light onto the physical description of this region. For future X-ray telescope missions this exercise may be trivial. The top panel of Figure \ref{fig:dec19_fig1} shows a simulated \textit{Athena-XIFU}\footnote{\textit{XIFU} response files used were XIFU\_CC\_BASELINECONF\_2018\_10\_10, retrieved from \url{http://x-ifu-resources.irap.omp.eu/PUBLIC/RESPONSES/CC_CONFIGURATION/}} 100\,\ks\ observation \citep{Barret+2022}, using the best fit \xstar+\apec\ model. Note that unlike with RGS, the \textit{XIFU} spectrum is several orders of magnitude   above the expected background. Most notably, \textit{XIFU} data will  distinguish between different ionizing continuum for the photoionized \xstar\ grids. The data were simulated with an ionizing continuum of $\Gamma=2.2$, equivalent to the continuum used for the majority of this work.  The four grids fit to the \textit{XIFU} data were ionized with: $\Gamma=2.2$; $\Gamma=2.5$; $\Gamma=1.5$; and an empirical continuum based on the best fit absorbed continuum  given by  \cite{Buhariwalla+2020}.  The $\dec$ value between the best fit $\Gamma=2.2$ grid and the $\Gamma=2.5$, $\Gamma=1.5$, and absorbed continuum were  $\dec=391$, $\dec=688$ and  $\dec=2476$, respectively.  Thus with \textit{XIFU} we  easily  place restrictions on the ionizing continuum for the photoionized material. This is not even taking into account that \textit{XIFU} will have high resolution spectroscopy up to $\sim$10 \kev\  allowing for direct testing of the link between the absorber and the photoionized emitter. 
%the  and to simulate the data

\subsection{The Collisional Plasma in Mrk 1239}
When testing spectral models no combination of photoionized emitters could fit the $13-17$\,\AA\ region well. Only the \apec\ and \mekal\ models could adequately fit the data in this region. Figure \ref{fig:aug8fig3} shows that this is the region of the RGS spectra that is dominated by the \apec\ component. The luminosity of each component was calculated  using the \xspec\ command  {\sc clumin}. The \apec\ component had a luminosity of  $L_{0.3-2.5 \kev}=41.02^{+0.06}_{-0.07}$ \ergps, the \xstar\ had $L_{0.3-2.5 \kev}=41.05^{+0.06}_{-0.08}$ \ergps. The brightness of each feature was approximately equal.

Emission from  \fexvii\ at  16.777\,\AA\  (3F), 17.050\,\AA\ (3G), and 17.097\,\AA\ (M2)  are commonly seen in collisionally ionized plasma. The 3G and M2 lines are  blended  due to instrument resolution. They are often labelled $I$(17.076\,\AA) as they appear as one emission line feature at that wavelength. The \ovii\ RRC is found at 16.769\,\AA\ which blends with the 3F line much the same way 3G and M2 blend ($I$(16.777\,\AA)). \cite{Phillips+1997} measured the ratios of 3G/3F and M2/3F lines in the spectrum of the solar corona. Using their data the  average value of ratio (3F)/(3G + M2) was calculated to be $0.48 \pm 0.2$. This suggests that if the  ratio of $I$(16.777\,\AA)/$I$(17.076\,\AA) is measured to be above 1 then than 50\% of the flux in the $I$(16.777\,\AA) feature is contributed by the \ovii\ RRC component. 

\cite{Bianchi+2010} found that the {\sc apec} model produced a ratio of $\approx 0.44$ at varying plasma temperatures. They then measured the $I$(16.777\,\AA)/$I$(17.076\,\AA) ratio in Mrk 573 to be $1.2_{-0.5}^{+0.7}$, concluding that the \ovii\ RRC feature was contaminated by the 3F emission line. 
Similarly \cite{Marinucci+2011} measured the ratio in  NGC 424 to be $2.6_{-1.9}^{+1.8}$, and stated that the RRC component dominated over the \fexvii\ lines. In \mrk, the ratio is measured at $0.9_{-0.4}^{+0.7}$ indicating that both the \fexvii\ emission and \ovii\ RRC are contributing to $I$(16.777\,\AA).

\cite{Guainazzi+2007} studied obscured AGN (Sy 1.5 or greater) and found  16 of the 69 objects they investigated  showed evidence of the \ovii\ RRC feature. They took a ratio of 0.6 to indicate the RRC feature was dominate of the 3F emission line. Of those 16 galaxies, 15 had a ratio >0.6, and 13 were >1.2.  \mrk\ ($0.9_{-0.4}^{+0.7}$) fits in the lower end of obscured AGN distribution and would be classified as having a dominate \ovii\ RRC feature over the 3F line by \cite{Guainazzi+2007}.

When the best-fit {\sc xstar} model was applied to the data without the {\sc apec} component and the (3F)/(3G +M2) ratio  was measure  it  resulted in a value <0.5. This indicated that the photoionized emission was accounting for the \ovii\ RRC feature and the remaining flux in this region was due to \fexvii\ emission from collisionally ionized material.  

Collisionally ionized emitters have been used to model the X-ray spectra of star forming regions in galaxies.  \cite{Franceschini+2003}  approximated the SFR from the  $L_{2-10 \kev}$ in ULIRGs without dominant AGN components. It is given by: 
\begin{equation}
{\rm SFR^{ULIRG}_{X-ray}} \approx \frac{L_{2-10 \kev}}{10^{39} {\rm erg\; s^{-1}}} \Msun {\rm yr^{-1}}\;.
\end{equation}
Using just the \apec\ component allows us to estimate the non-AGN contribution of the  2-10\kev\ luminosity. We measure the \apec\ contribution to be $L_{2-10 \kev} \approx \num{3e39}$ \ergps, resulting in a SFR of 3 $\Msun \, {\rm yr^{-1}}$.  This is consistent with  previous measurements SFR in \mrk\ such as X-ray measurements \citep[$ 3.7-5.8\Msun \, {\rm yr^{-1}}$;][]{Buhariwalla+2020},  PAH measurements \citep[$\lesssim 7.5\Msun \, {\rm yr^{-1}}$;][]{RuschelDutra+2017},    SED fitting \citep[$3.47\pm 0.26\Msun \, {\rm yr^{-1}}$;][]{Gruppioni+2016}, and IR measurements \citep[$2.1^{+0.5}_{-0.4}\Msun \, {\rm yr^{-1}}$;][]{Smirnova+2022} . In the sample by \cite{RuschelDutra+2017}, the luminosity from the collisionally ionized component matched those galaxies whose X-ray spectra  were dominated by starburst component rather than  those dominated by AGN. 

\subsection{Mrk\,1239 Compared to Seyfert Galaxies}
In a sample of 21 bright \rosat\ selected AGN  viewed with \xmm-\textit{EPIC}, only 2 of the galaxies studied preferred a soft excess with a line emitting region  over a smooth blackbody  \citep{Gallo+2006C}. The sample included 19 Seyfert galaxies (from Seyfert 1s to Seyfert 1.9s) and 2  LINERs and it was the two LINERs that required the CI component (\mekal).

The RGS spectrum of Seyfert 1 galaxies is typically dominated by features from  photoionized plasma. This primarily includes emission from He-like ion triplets (e.g. \ovii, \neix, N\,{\sc vi})  and RRC features (\oviii, \neix, C\,{\sc vi}).   NGC 5548 is among the first Seyfert galaxies studied \citep{Seyfert+1943}, it possesses  narrow and broad emission features in both the optical and X-ray bands \citep{Cappi+2016,Whewell+2015}. The RGS spectra of this object shows RRC features from \neix, \ovii, \oviii, C\,{\sc v}, and C\,{\sc vi}. In addition the spectra show a strong \ovii$(f)$ line compared to the $i$ and $r$ lines. This galaxy shows strong evidence of several photoionized emitters and no evidence of any collisionally ionized material \citep{Mao+2018}.   

NGC 4151 is another well known Seyfert 1.5 galaxy. It shows many of the same features that NGC 5548 does\ including RRC and He-like triplets. The $R$ and $G$ ratios straddle the boarder between photoionized and collisional ionized plasma. However, the strength and shape of the RRC features in this galaxy point to a purely photoionized plasma \citep{Armentrout+2007}.

The NLS1, Mrk\,335 has been extensively studied with \xmm. While early RGS spectra with low signal-to-noise supported a CI plasma \citep{Grupe+2008}, theses were not confirmed in later works (e.g. \citealt{Longinotti+2013,Longinotti+2019}). \cite{Parker+2019} examined the low continuum flux state when the emission lines were strongest and attributed them to photoionized emission. 

\mrk\ shows strong evidence of collisionally ionized plasma, putting it at odds with the many of Seyfert 1 observations that favour photoionized emission.  In \mrk\ the measured  ratio of the forbidden \ovii\ line to the \oviii\ Ly$_{\alpha}$ emission line is $\approx 1$ as is  expected for an obscured AGN (Sy 1.5 or greater) compared to the lower ratio for starburst galaxy \citep{Guainazzi+2007}. 

NGC 1068 is a prototypical Seyfert 2 galaxy, and it shows a myriad of emission features stemming from multiple photoionized components. Based on visual inspection this galaxy has a spectrum that looks similar to \mrk. The He-like triplets in NGC\,1068 have similar appearances, with the  \neix\ triplet showing equally strong $r$ and $f$ lines and no $i$ line matching \mrk\ completely \citep{Grafton+2021}. However the \neix\ triplet is not as blended with other lines in NGC 1068 as in \mrk. 

Original analysis of NGC\,1068 showed that the maximum strength of a collisionally ionized plasma would be an order of magnitude less than the photoionized plasma \citep{Kinkhabwala+2002}. However, the photoionization models used to describe the RGS spectrum failed to properly account for \fexvii\ at 15\,\AA\ and 17\,\AA, thus \cite{Grafton+2021} used a collisionally ionized component to model these lines. They stressed that this was most likely not caused by starburst activity. Instead it may be due to incomplete photoexcitation information in {\sc spex},  causing \pion\ to under predict emission in these lines.  As for \mrk\ this does not seem to be the case as no matter the number of photoionized components applied (in addition to powerlaw components) the spectra is still under fit between $13-17$\,\AA\ without the addition of \apec.

 \mrk\ is a polar scattered Seyfert 1 galaxy, meaning our line of sight (LOS) passes through the upper layers of the torus polarizing the optical emission    \citep{Smith+2004,Jiang+2021}. ESO 323-G77, another polar scattered Seyfert1,   presents an X-ray spectra similar  to \mrk. This object shows high levels of variable absorption, that leave the soft band remarkably consistent below 1.5 \kev\ (\mrk\ is consistent below 3 \kev) and the harder bands show  typical Seyfert variability (the same in \mrk). In the soft band of ESO 323-G77 excess emission is seen at 0.9 \kev, and can be modelled using a collisionally ionized plasma. This plasma has been interpreted as a region of star formation activity, and a SFR can be extracted in much the same way as is done here in \mrk\   \citep{Miniutti+2014}. 
 	
ESO 323-G77 has been interpreted to be obscured by a clumpy torus, BLR clouds and a warm outflowing medium  \citep{Miniutti+2014}. 
The soft X-ray  band of this object has been modelled using  two \apec\ components, one with similar temperature to \mrk\ ($kT=0.74$\kev) , the other much cooler, $kT=0.09$\kev. This is a difference in the modelling of these  two objects, as the  cool \apec\ component in ESO 323-G77 takes the place of a photoionized emitter. This cool \apec\ component fits  a slight excess emission around $\sim0.5$ \kev, however it could be argued that the this excess emission could be originating from photoionized plasma  (\ovii\ triplet/N\,{\sc vii}  emission lines), but the CCD spectra lacks the resolution to distinguish between the components. \mrk\ could have a geometry similar to this object  where  the LOS is through the torus, creating a polarized BLR and absorbing the X-ray continuum, allowing for distant star forming regions to be visible in the  soft X-ray band. ESO 323-G77  may exhibit a similar RGS spectrum as we have seen here in \mrk.

If the soft X-ray band is \mrk\ is obscured in a similar manor as it is in ESO 323-G77, then an interesting avenue for analysis would be to investigate the connections with the photoionized emitter producing the \ovii\ triplet and the obscuring torus. Both sources would benefit from much deeper observations to explore this connection. 
 
\section{Conclusions}
\label{sec:conclusion}
In this work we present the first deep RGS spectrum of \mrk, it  shows a myriad of ionized emission lines originating from a blend of photoionized  and collisionally ionized plasma. The collisionally ionized plasma dominates the RGS spectra below 17\,\AA, while the photoionized plasma dominates above 17\,\AA. The \ovii\ triplet is detected for the very first time in this source. Our main conclusions are as follows:
\begin{itemize}
	\item[i.] Based on a strong 0.91 \kev\ feature interpreted as \neix\ triplet and the lack of \ovii\ triplet,  \cite{Grupe+2004} concluded that there was a super solar abundance of  Ne/O. We show that the \neix\ triplet is enhanced in a CCD spectra  by Fe-L transitions, as proposed by \cite{Buhariwalla+2020}, and that the \ovii$(f)$ triplet is present. All fits done in this work use a solar abundance of Ne/O, allowing us to conclude that \mrk\ does not have an over abundance of Ne. 
	\item[ii.]  Based on the line ratios of the \ovii\ triplet and the variable density \xstar\ grids, we can infer that the density of the photoionized material is between $10^5-10^{10} \cm^{-3}$. Using the definition of ionization and luminosities  estimated from \cite{Buhariwalla+2020} we can estimate that the photoionized emitter is located within a few \pc\ of the SMBH. Based in on the FWHM of the \ovii$(f)$ emission line, the emitting material cannot be closer than the inner radius of the torus. 
	\item[iii.] The collisionally ionized material is interpenetrated as a region of star forming activity. Based on relations from \cite{Franceschini+2003} we estimate the SFR to be $\sim3 \Msun \, {\rm yr^{-1}}$. This is consistent with the the SFR found by \cite{Buhariwalla+2020} using the same method, and using PAH measurement of \mrk\ \citep{RuschelDutra+2017}.   
	\item[iv.] Seyfert 1 galaxies typically show evidence of photoionized emission, it is uncommon for them to show evidence of collisionally ionized plasma in their X-ray spectra, placing \mrk\ in the minority along with ESO 323-G77. 
\end{itemize}
The broadband spectrum \mrk\ will be presented in a future work. We will investigate any connection between the photoionized emitter and the highly absorbed continuum. 
\section*{Acknowledgements}
This work was based on observations obtained with XMM-Newton, an ESA science mission with instruments and contributions directly funded by ESA Member States and NASA. This research has also made use of data obtained from the Chandra Data Archive and the Chandra Source Catalog, and software provided by the Chandra X-ray Center (CXC) in the application packages CIAO and Sherpa.
We thank the referee for comments that helped clarify the work. LCG acknowledges financial support from the Natural Sciences and Engineering Research Council of Canada (NSERC) and from the Canadian Space Agency (CSA). JJ acknowledges support from the Leverhulme Trust, Isaac Newton Trust, and St. Edmund's College University of Cambridge. 

%%%%%%%%%%%%%%%%%%%%%%%%%%%%%%%%%%%%%%%%%%%%%%%%%%
\section*{Data Availability}
All data will be available through the HEASARC  archive after the proprietary period has ended.


\bibliographystyle{mnras}
\bibliography{bibtext} % if your bibtex file is called example.bib

\begin{thebibliography}{}
\makeatletter
\relax
\def\mn@urlcharsother{\let\do\@makeother \do\$\do\&\do\#\do\^\do\_\do\%\do\~}
\def\mn@doi{\begingroup\mn@urlcharsother \@ifnextchar [ {\mn@doi@}
  {\mn@doi@[]}}
\def\mn@doi@[#1]#2{\def\@tempa{#1}\ifx\@tempa\@empty \href
  {http://dx.doi.org/#2} {doi:#2}\else \href {http://dx.doi.org/#2} {#1}\fi
  \endgroup}
\def\mn@eprint#1#2{\mn@eprint@#1:#2::\@nil}
\def\mn@eprint@arXiv#1{\href {http://arxiv.org/abs/#1} {{\tt arXiv:#1}}}
\def\mn@eprint@dblp#1{\href {http://dblp.uni-trier.de/rec/bibtex/#1.xml}
  {dblp:#1}}
\def\mn@eprint@#1:#2:#3:#4\@nil{\def\@tempa {#1}\def\@tempb {#2}\def\@tempc
  {#3}\ifx \@tempc \@empty \let \@tempc \@tempb \let \@tempb \@tempa \fi \ifx
  \@tempb \@empty \def\@tempb {arXiv}\fi \@ifundefined
  {mn@eprint@\@tempb}{\@tempb:\@tempc}{\expandafter \expandafter \csname
  mn@eprint@\@tempb\endcsname \expandafter{\@tempc}}}

\bibitem[\protect\citeauthoryear{{Arav}, {Barlow}, {Laor}, {Sargent}  \&
  {Blandford}}{{Arav} et~al.}{1998}]{Arav+1998}
{Arav} N.,  {Barlow} T.~A.,  {Laor} A.,  {Sargent} W. L.~W.,   {Blandford}
  R.~D.,  1998, \mn@doi [\mnras] {10.1046/j.1365-8711.1998.297004990.x}, \href
  {https://ui.adsabs.harvard.edu/abs/1998MNRAS.297..990A} {297, 990}

\bibitem[\protect\citeauthoryear{{Armentrout}, {Kraemer}  \&
  {Turner}}{{Armentrout} et~al.}{2007}]{Armentrout+2007}
{Armentrout} B.~K.,  {Kraemer} S.~B.,   {Turner} T.~J.,  2007, \mn@doi [\apj]
  {10.1086/519512}, \href
  {https://ui.adsabs.harvard.edu/abs/2007ApJ...665..237A} {665, 237}

\bibitem[\protect\citeauthoryear{{Barret} et~al.,}{{Barret}
  et~al.}{2022}]{Barret+2022}
{Barret} D.,  et~al., 2022, arXiv e-prints, \href
  {https://ui.adsabs.harvard.edu/abs/2022arXiv220814562B} {p. arXiv:2208.14562}

\bibitem[\protect\citeauthoryear{{Beers}, {Kriessler}, {Bird}  \&
  {Huchra}}{{Beers} et~al.}{1995}]{Beers+1995}
{Beers} T.~C.,  {Kriessler} J.~R.,  {Bird} C.~M.,   {Huchra} J.~P.,  1995,
  \mn@doi [\aj] {10.1086/117329}, \href
  {https://ui.adsabs.harvard.edu/abs/1995AJ....109..874B} {109, 874}

\bibitem[\protect\citeauthoryear{{Bennert}, {Falcke}, {Schulz}, {Wilson}  \&
  {Wills}}{{Bennert} et~al.}{2002}]{Bennert+2002}
{Bennert} N.,  {Falcke} H.,  {Schulz} H.,  {Wilson} A.~S.,   {Wills} B.~J.,
  2002, \mn@doi [\apjl] {10.1086/342420}, \href
  {https://ui.adsabs.harvard.edu/abs/2002ApJ...574L.105B} {574, L105}

\bibitem[\protect\citeauthoryear{{Bennert}, {Jungwiert}, {Komossa}, {Haas}  \&
  {Chini}}{{Bennert} et~al.}{2006}]{Bennert+2006}
{Bennert} N.,  {Jungwiert} B.,  {Komossa} S.,  {Haas} M.,   {Chini} R.,  2006,
  \mn@doi [\aap] {10.1051/0004-6361:20065319}, \href
  {https://ui.adsabs.harvard.edu/abs/2006A&A...456..953B} {456, 953}

\bibitem[\protect\citeauthoryear{{Bianchi}, {Chiaberge}, {Evans}, {Guainazzi},
  {Baldi}, {Matt}  \& {Piconcelli}}{{Bianchi} et~al.}{2010}]{Bianchi+2010}
{Bianchi} S.,  {Chiaberge} M.,  {Evans} D.~A.,  {Guainazzi} M.,  {Baldi} R.~D.,
   {Matt} G.,   {Piconcelli} E.,  2010, \mn@doi [\mnras]
  {10.1111/j.1365-2966.2010.16475.x}, \href
  {https://ui.adsabs.harvard.edu/abs/2010MNRAS.405..553B} {405, 553}

\bibitem[\protect\citeauthoryear{{Buhariwalla}, {Waddell}, {Gallo}, {Grupe}  \&
  {Komossa}}{{Buhariwalla} et~al.}{2020}]{Buhariwalla+2020}
{Buhariwalla} M.~Z.,  {Waddell} S. G.~H.,  {Gallo} L.~C.,  {Grupe} D.,
  {Komossa} S.,  2020, \mn@doi [\apj] {10.3847/1538-4357/abb08a}, \href
  {https://ui.adsabs.harvard.edu/abs/2020ApJ...901..118B} {901, 118}

\bibitem[\protect\citeauthoryear{{Cappi} et~al.,}{{Cappi}
  et~al.}{2016}]{Cappi+2016}
{Cappi} M.,  et~al., 2016, \mn@doi [\aap] {10.1051/0004-6361/201628464}, \href
  {https://ui.adsabs.harvard.edu/abs/2016A&A...592A..27C} {592, A27}

\bibitem[\protect\citeauthoryear{{Cash}}{{Cash}}{1979}]{Cash+1979}
{Cash} W.,  1979, \mn@doi [ApJ] {10.1086/156922}, \href
  {https://ui.adsabs.harvard.edu/abs/1979ApJ...228..939C} {228, 939}

\bibitem[\protect\citeauthoryear{{Doi}, {Wajima}, {Hagiwara}  \& {Inoue}}{{Doi}
  et~al.}{2015}]{Doi+2015}
{Doi} A.,  {Wajima} K.,  {Hagiwara} Y.,   {Inoue} M.,  2015, \mn@doi [\apjl]
  {10.1088/2041-8205/798/2/L30}, \href
  {https://ui.adsabs.harvard.edu/abs/2015ApJ...798L..30D} {798, L30}

\bibitem[\protect\citeauthoryear{{Foord}}{{Foord}}{2020}]{Foord+2020}
{Foord} A.,  2020, {Dual AGN Across Cosmic Time}, Chandra Proposal ID
  \#22700153

\bibitem[\protect\citeauthoryear{{Franceschini} et~al.,}{{Franceschini}
  et~al.}{2003}]{Franceschini+2003}
{Franceschini} A.,  et~al., 2003, \mn@doi [\mnras]
  {10.1046/j.1365-8711.2003.06744.x}, \href
  {https://ui.adsabs.harvard.edu/abs/2003MNRAS.343.1181F} {343, 1181}

\bibitem[\protect\citeauthoryear{Gallo}{Gallo}{2018}]{Gallo+2018}
Gallo L.,  2018, \mn@doi [PoS] {10.22323/1.328.0034}, NLS1-2018, 034

\bibitem[\protect\citeauthoryear{{Gallo}, {Lehmann}, {Pietsch}, {Boller},
  {Brinkmann}, {Friedrich}  \& {Grupe}}{{Gallo} et~al.}{2006}]{Gallo+2006C}
{Gallo} L.~C.,  {Lehmann} I.,  {Pietsch} W.,  {Boller} T.,  {Brinkmann} W.,
  {Friedrich} P.,   {Grupe} D.,  2006, \mn@doi [\mnras]
  {10.1111/j.1365-2966.2005.09755.x}, \href
  {https://ui.adsabs.harvard.edu/abs/2006MNRAS.365..688G} {365, 688}

\bibitem[\protect\citeauthoryear{{Goodrich}}{{Goodrich}}{1989}]{Goodrich+1989}
{Goodrich} R.~W.,  1989, \mn@doi [ApJ] {10.1086/167586}, \href
  {https://ui.adsabs.harvard.edu/abs/1989ApJ...342..224G} {342, 224}

\bibitem[\protect\citeauthoryear{{Grafton-Waters}, {Branduardi-Raymont},
  {Mehdipour}, {Page}, {Bianchi}, {Behar}  \& {Symeonidis}}{{Grafton-Waters}
  et~al.}{2021}]{Grafton+2021}
{Grafton-Waters} S.,  {Branduardi-Raymont} G.,  {Mehdipour} M.,  {Page} M.,
  {Bianchi} S.,  {Behar} E.,   {Symeonidis} M.,  2021, \mn@doi [\aap]
  {10.1051/0004-6361/202039022}, \href
  {https://ui.adsabs.harvard.edu/abs/2021A&A...649A.162G} {649, A162}

\bibitem[\protect\citeauthoryear{Grupe, Mathur  \& Komossa}{Grupe
  et~al.}{2004}]{Grupe+2004}
Grupe D.,  Mathur S.,   Komossa S.,  2004, \mn@doi [\aj] {10.1086/421002}, 127,
  3161

\bibitem[\protect\citeauthoryear{Grupe, Komossa, Gallo, Fabian, Larsson,
  Pradhan, Xu  \& Miniutti}{Grupe et~al.}{2008}]{Grupe+2008}
Grupe D.,  Komossa S.,  Gallo L.~C.,  Fabian A.~C.,  Larsson J.,  Pradhan
  A.~K.,  Xu D.,   Miniutti G.,  2008, \mn@doi [The Astrophysical Journal]
  {10.1086/588213}, 681, 982

\bibitem[\protect\citeauthoryear{{Gruppioni} et~al.,}{{Gruppioni}
  et~al.}{2016}]{Gruppioni+2016}
{Gruppioni} C.,  et~al., 2016, \mn@doi [\mnras] {10.1093/mnras/stw577}, \href
  {https://ui.adsabs.harvard.edu/abs/2016MNRAS.458.4297G} {458, 4297}

\bibitem[\protect\citeauthoryear{{Guainazzi} \& {Bianchi}}{{Guainazzi} \&
  {Bianchi}}{2007}]{Guainazzi+2007}
{Guainazzi} M.,  {Bianchi} S.,  2007, \mn@doi [\mnras]
  {10.1111/j.1365-2966.2006.11229.x}, \href
  {https://ui.adsabs.harvard.edu/abs/2007MNRAS.374.1290G} {374, 1290}

\bibitem[\protect\citeauthoryear{{Husemann} et~al.,}{{Husemann}
  et~al.}{2022}]{Husemann+2022}
{Husemann} B.,  et~al., 2022, \mn@doi [\aap] {10.1051/0004-6361/202141312},
  \href {https://ui.adsabs.harvard.edu/abs/2022A&A...659A.124H} {659, A124}

\bibitem[\protect\citeauthoryear{{Jansen} et~al.,}{{Jansen}
  et~al.}{2001}]{Jansen+2001}
{Jansen} F.,  et~al., 2001, \mn@doi [AAP] {10.1051/0004-6361:20000036}, \href
  {https://ui.adsabs.harvard.edu/abs/2001A&A...365L...1J} {365, L1}

\bibitem[\protect\citeauthoryear{{J{\"a}rvel{\"a}}, {Dahale}, {Crepaldi},
  {Berton}, {Congiu}  \& {Antonucci}}{{J{\"a}rvel{\"a}}
  et~al.}{2022}]{Jarvela+2022}
{J{\"a}rvel{\"a}} E.,  {Dahale} R.,  {Crepaldi} L.,  {Berton} M.,  {Congiu} E.,
    {Antonucci} R.,  2022, \mn@doi [\aap] {10.1051/0004-6361/202141698}, \href
  {https://ui.adsabs.harvard.edu/abs/2022A&A...658A..12J} {658, A12}

\bibitem[\protect\citeauthoryear{{Jiang} et~al.,}{{Jiang}
  et~al.}{2019}]{Jiang+2019}
{Jiang} J.,  et~al., 2019, \mn@doi [\mnras] {10.1093/mnras/stz2326}, \href
  {https://ui.adsabs.harvard.edu/abs/2019MNRAS.489.3436J} {489, 3436}

\bibitem[\protect\citeauthoryear{{Jiang}, {Balokovi{\'c}}, {Brightman}, {Liu},
  {Harrison}  \& {Lansbury}}{{Jiang} et~al.}{2021}]{Jiang+2021}
{Jiang} J.,  {Balokovi{\'c}} M.,  {Brightman} M.,  {Liu} H.,  {Harrison} F.~A.,
    {Lansbury} G.~B.,  2021, \mn@doi [\mnras] {10.1093/mnras/stab1306}, \href
  {https://ui.adsabs.harvard.edu/abs/2021MNRAS.505..702J} {505, 702}

\bibitem[\protect\citeauthoryear{{Kaastra} \& {Bleeker}}{{Kaastra} \&
  {Bleeker}}{2016}]{optbin}
{Kaastra} J.~S.,  {Bleeker} J.~A.~M.,  2016, \mn@doi [AAP]
  {10.1051/0004-6361/201527395}, \href
  {https://ui.adsabs.harvard.edu/abs/2016A&A...587A.151K} {587, A151}

\bibitem[\protect\citeauthoryear{{Kallman} \& {Bautista}}{{Kallman} \&
  {Bautista}}{2001}]{Kallman+2001}
{Kallman} T.,  {Bautista} M.,  2001, \mn@doi [\apjs] {10.1086/319184}, \href
  {https://ui.adsabs.harvard.edu/abs/2001ApJS..133..221K} {133, 221}

\bibitem[\protect\citeauthoryear{{Kaspi}, {Maoz}, {Netzer}, {Peterson},
  {Vestergaard}  \& {Jannuzi}}{{Kaspi} et~al.}{2005}]{Kaspi+2005}
{Kaspi} S.,  {Maoz} D.,  {Netzer} H.,  {Peterson} B.~M.,  {Vestergaard} M.,
  {Jannuzi} B.~T.,  2005, \mn@doi [\apj] {10.1086/431275}, \href
  {https://ui.adsabs.harvard.edu/abs/2005ApJ...629...61K} {629, 61}

\bibitem[\protect\citeauthoryear{{Kinkhabwala} et~al.,}{{Kinkhabwala}
  et~al.}{2002}]{Kinkhabwala+2002}
{Kinkhabwala} A.,  et~al., 2002, \mn@doi [\apj] {10.1086/341482}, \href
  {https://ui.adsabs.harvard.edu/abs/2002ApJ...575..732K} {575, 732}

\bibitem[\protect\citeauthoryear{{Komossa}}{{Komossa}}{2008}]{Komossa+2008}
{Komossa} S.,  2008, in Rev. Mex. Astron. Astrofis. Conference Series. pp
  86--92 (\mn@eprint {arXiv} {0710.3326})

\bibitem[\protect\citeauthoryear{{Komossa} \& {Schulz}}{{Komossa} \&
  {Schulz}}{1997}]{Komossa+1997}
{Komossa} S.,  {Schulz} H.,  1997, \aap, \href
  {https://ui.adsabs.harvard.edu/abs/1997A&A...323...31K} {323, 31}

\bibitem[\protect\citeauthoryear{{Liedahl}, {Osterheld}  \&
  {Goldstein}}{{Liedahl} et~al.}{1995}]{Liedahl+1995}
{Liedahl} D.~A.,  {Osterheld} A.~L.,   {Goldstein} W.~H.,  1995, \mn@doi
  [\apjl] {10.1086/187729}, \href
  {https://ui.adsabs.harvard.edu/abs/1995ApJ...438L.115L} {438, L115}

\bibitem[\protect\citeauthoryear{{Longinotti} et~al.,}{{Longinotti}
  et~al.}{2013}]{Longinotti+2013}
{Longinotti} A.~L.,  et~al., 2013, \mn@doi [\apj]
  {10.1088/0004-637X/766/2/104}, \href
  {https://ui.adsabs.harvard.edu/abs/2013ApJ...766..104L} {766, 104}

\bibitem[\protect\citeauthoryear{{Longinotti} et~al.,}{{Longinotti}
  et~al.}{2019}]{Longinotti+2019}
{Longinotti} A.~L.,  et~al., 2019, \mn@doi [\apj] {10.3847/1538-4357/ab125a},
  \href {https://ui.adsabs.harvard.edu/abs/2019ApJ...875..150L} {875, 150}

\bibitem[\protect\citeauthoryear{{Malkan}, {Jensen}, {Rodriguez}, {Spinoglio}
  \& {Rush}}{{Malkan} et~al.}{2017}]{Malkan+2017}
{Malkan} M.~A.,  {Jensen} L.~D.,  {Rodriguez} D.~R.,  {Spinoglio} L.,   {Rush}
  B.,  2017, \mn@doi [\apj] {10.3847/1538-4357/aa8302}, \href
  {https://ui.adsabs.harvard.edu/abs/2017ApJ...846..102M} {846, 102}

\bibitem[\protect\citeauthoryear{{Mao} et~al.,}{{Mao} et~al.}{2018}]{Mao+2018}
{Mao} J.,  et~al., 2018, \mn@doi [\aap] {10.1051/0004-6361/201732162}, \href
  {https://ui.adsabs.harvard.edu/abs/2018A&A...612A..18M} {612, A18}

\bibitem[\protect\citeauthoryear{{Mao} et~al.,}{{Mao} et~al.}{2019}]{Mao+2019}
{Mao} J.,  et~al., 2019, \mn@doi [\aap] {10.1051/0004-6361/201730931}, \href
  {https://ui.adsabs.harvard.edu/abs/2019A&A...621A...9M} {621, A9}

\bibitem[\protect\citeauthoryear{{Marinucci}, {Bianchi}, {Matt}, {Fabian},
  {Iwasawa}, {Miniutti}  \& {Piconcelli}}{{Marinucci}
  et~al.}{2011}]{Marinucci+2011}
{Marinucci} A.,  {Bianchi} S.,  {Matt} G.,  {Fabian} A.~C.,  {Iwasawa} K.,
  {Miniutti} G.,   {Piconcelli} E.,  2011, \mn@doi [\aap]
  {10.1051/0004-6361/201015358}, \href
  {https://ui.adsabs.harvard.edu/abs/2011A&A...526A..36M} {526, A36}

\bibitem[\protect\citeauthoryear{{Miniutti} et~al.,}{{Miniutti}
  et~al.}{2014}]{Miniutti+2014}
{Miniutti} G.,  et~al., 2014, \mn@doi [\mnras] {10.1093/mnras/stt2005}, \href
  {https://ui.adsabs.harvard.edu/abs/2014MNRAS.437.1776M} {437, 1776}

\bibitem[\protect\citeauthoryear{Netzer}{Netzer}{2013}]{Netzer+2013}
Netzer H.,  2013, The Physics and Evolution of Active Galactic Nuclei..
Cambridge University Press, \url
  {https://search.ebscohost.com/login.aspx?direct=true&db=nlebk&AN=527870&site=ehost-live}

\bibitem[\protect\citeauthoryear{{Osterbrock} \& {Pogge}}{{Osterbrock} \&
  {Pogge}}{1985}]{Osterbrock+1985}
{Osterbrock} D.~E.,  {Pogge} R.~W.,  1985, \mn@doi [ApJ] {10.1086/163513},
  \href {https://ui.adsabs.harvard.edu/abs/1985ApJ...297..166O} {297, 166}

\bibitem[\protect\citeauthoryear{{Pan} et~al.,}{{Pan} et~al.}{2019}]{Pan+2019}
{Pan} X.,  et~al., 2019, \mn@doi [\apj] {10.3847/1538-4357/aaf1bc}, \href
  {https://ui.adsabs.harvard.edu/abs/2019ApJ...870...75P} {870, 75}

\bibitem[\protect\citeauthoryear{{Pan} et~al.,}{{Pan} et~al.}{2021}]{Pan+2021}
{Pan} X.,  et~al., 2021, \mn@doi [\apj] {10.3847/1538-4357/abf148}, \href
  {https://ui.adsabs.harvard.edu/abs/2021ApJ...912..118P} {912, 118}

\bibitem[\protect\citeauthoryear{Parker et~al.,}{Parker
  et~al.}{2019}]{Parker+2019}
Parker M.~L.,  et~al., 2019, \mn@doi [MNRAS] {10.1093/mnras/stz2566}, 490, 683

\bibitem[\protect\citeauthoryear{{Phillips}, {Greer}, {Bhatia}, {Coffey},
  {Barnsley}  \& {Keenan}}{{Phillips} et~al.}{1997}]{Phillips+1997}
{Phillips} K.~J.~H.,  {Greer} C.~J.,  {Bhatia} A.~K.,  {Coffey} I.~H.,
  {Barnsley} R.,   {Keenan} F.~P.,  1997, \aap, \href
  {https://ui.adsabs.harvard.edu/abs/1997A&A...324..381P} {324, 381}

\bibitem[\protect\citeauthoryear{{Porquet} \& {Dubau}}{{Porquet} \&
  {Dubau}}{2000}]{Porquet+2000}
{Porquet} D.,  {Dubau} J.,  2000, \mn@doi [\aaps] {10.1051/aas:2000192}, \href
  {https://ui.adsabs.harvard.edu/abs/2000A&AS..143..495P} {143, 495}

\bibitem[\protect\citeauthoryear{{Rafanelli} \& {Bonoli}}{{Rafanelli} \&
  {Bonoli}}{1984}]{Rafanelli+1984}
{Rafanelli} P.,  {Bonoli} C.,  1984, \aap, \href
  {https://ui.adsabs.harvard.edu/abs/1984A&A...131..186R} {131, 186}

\bibitem[\protect\citeauthoryear{Ruschel-Dutra, Rodr\'{i}guez~Espinosa,
  Gonz\'{a}lez~Mart\'{a}n, Pastoriza  \& Riffel}{Ruschel-Dutra
  et~al.}{2016}]{RuschelDutra+2017}
Ruschel-Dutra D.,  Rodr\'{i}guez~Espinosa J.~M.,  Gonz\'{a}lez~Mart\'{a}n O.,
  Pastoriza M.,   Riffel R.,  2016, \mn@doi [MNRAS] {10.1093/mnras/stw3276},
  466, 3353

\bibitem[\protect\citeauthoryear{{Rush} \& {Malkan}}{{Rush} \&
  {Malkan}}{1996}]{Rush+1996}
{Rush} B.,  {Malkan} M.~A.,  1996, \mn@doi [\apj] {10.1086/176672}, \href
  {https://ui.adsabs.harvard.edu/abs/1996ApJ...456..466R} {456, 466}

\bibitem[\protect\citeauthoryear{{Schmitt}, {Donley}, {Antonucci}, {Hutchings},
  {Kinney}  \& {Pringle}}{{Schmitt} et~al.}{2003}]{Schmitt+2003}
{Schmitt} H.~R.,  {Donley} J.~L.,  {Antonucci} R.~R.~J.,  {Hutchings} J.~B.,
  {Kinney} A.~L.,   {Pringle} J.~E.,  2003, \mn@doi [\apj] {10.1086/381224},
  \href {https://ui.adsabs.harvard.edu/abs/2003ApJ...597..768S} {597, 768}

\bibitem[\protect\citeauthoryear{{Seyfert}}{{Seyfert}}{1943}]{Seyfert+1943}
{Seyfert} C.~K.,  1943, \mn@doi [\apj] {10.1086/144488}, \href
  {https://ui.adsabs.harvard.edu/abs/1943ApJ....97...28S} {97, 28}

\bibitem[\protect\citeauthoryear{{Smirnova-Pinchukova}
  et~al.,}{{Smirnova-Pinchukova} et~al.}{2022}]{Smirnova+2022}
{Smirnova-Pinchukova} I.,  et~al., 2022, \mn@doi [\aap]
  {10.1051/0004-6361/202142011}, \href
  {https://ui.adsabs.harvard.edu/abs/2022A&A...659A.125S} {659, A125}

\bibitem[\protect\citeauthoryear{{Smith}, {Brickhouse}, {Liedahl}  \&
  {Raymond}}{{Smith} et~al.}{2001}]{Smith+2001}
{Smith} R.~K.,  {Brickhouse} N.~S.,  {Liedahl} D.~A.,   {Raymond} J.~C.,  2001,
  \mn@doi [\apjl] {10.1086/322992}, \href
  {https://ui.adsabs.harvard.edu/abs/2001ApJ...556L..91S} {556, L91}

\bibitem[\protect\citeauthoryear{{Smith}, {Robinson}, {Alexander}, {Young},
  {Axon}  \& {Corbett}}{{Smith} et~al.}{2004}]{Smith+2004}
{Smith} J.~E.,  {Robinson} A.,  {Alexander} D.~M.,  {Young} S.,  {Axon} D.~J.,
   {Corbett} E.~A.,  2004, \mn@doi [\mnras] {10.1111/j.1365-2966.2004.07610.x},
  \href {https://ui.adsabs.harvard.edu/abs/2004MNRAS.350..140S} {350, 140}

\bibitem[\protect\citeauthoryear{{Tarter}, {Tucker}  \& {Salpeter}}{{Tarter}
  et~al.}{1969}]{Tarter+1969}
{Tarter} C.~B.,  {Tucker} W.~H.,   {Salpeter} E.~E.,  1969, \mn@doi [\apj]
  {10.1086/150026}, \href
  {https://ui.adsabs.harvard.edu/abs/1969ApJ...156..943T} {156, 943}

\bibitem[\protect\citeauthoryear{{Tristram} \& {Schartmann}}{{Tristram} \&
  {Schartmann}}{2011}]{Tristram+2011}
{Tristram} K.~R.~W.,  {Schartmann} M.,  2011, \mn@doi [\aap]
  {10.1051/0004-6361/201116867}, \href
  {https://ui.adsabs.harvard.edu/abs/2011A&A...531A..99T} {531, A99}

\bibitem[\protect\citeauthoryear{{Vaona}, {Ciroi}, {Di Mille}, {Cracco}, {La
  Mura}  \& {Rafanelli}}{{Vaona} et~al.}{2012}]{Vaona+2012}
{Vaona} L.,  {Ciroi} S.,  {Di Mille} F.,  {Cracco} V.,  {La Mura} G.,
  {Rafanelli} P.,  2012, \mn@doi [\mnras] {10.1111/j.1365-2966.2012.22060.x},
  \href {https://ui.adsabs.harvard.edu/abs/2012MNRAS.427.1266V} {427, 1266}

\bibitem[\protect\citeauthoryear{{Waddell} \& {Gallo}}{{Waddell} \&
  {Gallo}}{2020}]{Waddell+2020}
{Waddell} S.~G.~H.,  {Gallo} L.~C.,  2020, \mn@doi [\mnras]
  {10.1093/mnras/staa2783}, \href
  {https://ui.adsabs.harvard.edu/abs/2020MNRAS.498.5207W} {498, 5207}

\bibitem[\protect\citeauthoryear{{Waddell}, {Gallo}, {Gonzalez}, {Tripathi}  \&
  {Zoghbi}}{{Waddell} et~al.}{2019}]{Waddell+2019}
{Waddell} S.~G.~H.,  {Gallo} L.~C.,  {Gonzalez} A.~G.,  {Tripathi} S.,
  {Zoghbi} A.,  2019, \mn@doi [\mnras] {10.1093/mnras/stz2518}, \href
  {https://ui.adsabs.harvard.edu/abs/2019MNRAS.489.5398W} {489, 5398}

\bibitem[\protect\citeauthoryear{{Whewell} et~al.,}{{Whewell}
  et~al.}{2015}]{Whewell+2015}
{Whewell} M.,  et~al., 2015, \mn@doi [\aap] {10.1051/0004-6361/201526742},
  \href {https://ui.adsabs.harvard.edu/abs/2015A&A...581A..79W} {581, A79}

\bibitem[\protect\citeauthoryear{{Wilkins} \& {Gallo}}{{Wilkins} \&
  {Gallo}}{2015}]{Wilkins+2015B}
{Wilkins} D.~R.,  {Gallo} L.~C.,  2015, \mn@doi [\mnras]
  {10.1093/mnras/stv162}, \href
  {https://ui.adsabs.harvard.edu/abs/2015MNRAS.449..129W} {449, 129}

\bibitem[\protect\citeauthoryear{{Wilkins}, {Gallo}, {Silva}, {Costantini},
  {Brandt}  \& {Kriss}}{{Wilkins} et~al.}{2017}]{Wilkins+2017}
{Wilkins} D.~R.,  {Gallo} L.~C.,  {Silva} C.~V.,  {Costantini} E.,  {Brandt}
  W.~N.,   {Kriss} G.~A.,  2017, \mn@doi [\mnras] {10.1093/mnras/stx1814},
  \href {https://ui.adsabs.harvard.edu/abs/2017MNRAS.471.4436W} {471, 4436}

\bibitem[\protect\citeauthoryear{{Willingale}, {Starling}, {Beardmore},
  {Tanvir}  \& {O'Brien}}{{Willingale} et~al.}{2013}]{Willingale+2013}
{Willingale} R.,  {Starling} R.~L.~C.,  {Beardmore} A.~P.,  {Tanvir} N.~R.,
  {O'Brien} P.~T.,  2013, \mn@doi [\mnras] {10.1093/mnras/stt175}, \href
  {https://ui.adsabs.harvard.edu/abs/2013MNRAS.431..394W} {431, 394}

\makeatother
\end{thebibliography}

%%%%%%%%%%%%%%%%%%%%%%%%%%%%%%%%%%%%%%%%%%%%%%%%%%

%%%%%%%%%%%%%%%%% APPENDICES %%%%%%%%%%%%%%%%%%%%%

%%%%%%%%%%%%%%%%%%%%%%%%%%%%%%%%%%%%%%%%%%%%%%%%%%

% Don't change these lines
\bsp	% typesetting comment
\label{lastpage}
\end{document}